\documentclass[aps,pra,showpacs]{revtex4}
\usepackage{latexsym,amsbsy,graphicx}

\begin{document}

%\preprint{}

\title{Beyond the simple Proximity Force Approximation: geometrical effects
on the non-retarded Casimir interaction.}

% repeat the \author ..  \affiliation etc.  as needed
% author.  Explanatory text should go in the []'s, actual e-mail

% \affiliation command.  The \affiliation command should follow the
\author{Bo E. Sernelius and C E Rom\'{a}n-Vel\'{a}zquez}
\email[]{bos@ifm.liu.se}
\homepage[]{www.ifm.liu.se/~boser}
%\thanks{}
\affiliation{Department of Physics, Chemistry and Biology, Link\"{o}ping
University, SE-581 83 Link\"{o}ping, Sweden}

%option in \documentclass).  \noaffiliation is required (may also be

\date{\today}

\begin{abstract}
We study the geometrical corrections to the simple Proximity Force
Approximation for the non-retarded Casimir force.  We present analytical results for
the force between objects of various shapes and substrates, and between
pairs of objects.  We compare the results to those from more exact numerical calculations.  We
treat spheres, spheroids, cylinders, cubes, cones, and wings; the analytical PFA results
together with the geometrical correction factors are summarized in a table.
\end{abstract}

\pacs{68.90.+g, 71.10.-w, 71.36.+c, 41.20.-q, 03.50.De, 03.70.+k, 05.40.-a}
%\keywords{}

\maketitle

\section{Introduction}
%\subsection{}
%\subsubsection{}

J. D. van der Waals found empirically in 1873 that there is an attractive
force between non-polar atoms.  It took a long time, until 1930, before
there was an explanation to this force.  London \cite {Lond} gave the
explanation in terms of fluctuations in the electron density
within the atoms (fluctuating dipoles).  In an alternative description \cite{Ser} one
may, instead of discussing the particles, focus on the electromagnetic
fields.  The force may be expressed as a result of changes in the
zero-point energy of the electromagnetic normal modes of the system.  There
are modes associated with the atoms and modes associated with the vacuum. 
Casimir studied a more pure and idealized system consisting of two
perfectly reflecting metal plates.  In this geometry there are no modes
associated with the plates themselves; there are only vacuum modes.  The
presence of the plates changes the vacuum modes and their zero-point
energy.  He published his findings in a classical paper
\cite {Casi} in 1948, the Casimir force was born, 60 years ago.  Casimir's paper is one
of the most important papers in the history of physics since it
demonstrates that the boundary conditions of a system may change its
zero-point energy and hence its properties.  In the case of objects made
from real materials both the modes associated with the objects and the
vacuum modes contribute to the force \cite{Ser}.  The first type dominates
in the van der Waals region which is for smaller separations.  For larger
separations between the objects the vacuum modes dominate and the result is
the Casimir force.  In the idealized case of perfect metals there is no van
der Waals range.  In 1997 Lamoreaux \cite{Lamo} performed the first modern
high-precision measurements of the Casimir force.  The accuracy was good
enough to make the direct comparison with theory feasible.  This spurred a
burst of renewed interest in the Casimir effect.  

The dispersion forces (van der Waals and Casimir forces) decrease in size with separation much faster than the Coulomb force and the gravitational force do. This means that for macroscopic objects at macroscopic separations the dispersion forces are very weak compared to other forces. However, in the micrometer and nanometer ranges they often dominate. Also in biological systems they are very important; the Coulomb forces are screened by the abundance of mobile ions; the dispersion forces are not screened.

In the rapidly emerging field of nano-technology the dispersion forces have come to play an important role and one tries to exploit these forces in nano-mechanical devices. In the design of such devices it is important to develop precise
calculations of the Casimir forces between objects with different
geometries and in different configurations. It may be complicated to
calculate the force between objects of general shapes.  The van der Waals
and Casimir forces between half spaces is, however, not very difficult to
calculate.  For finite objects at large distances one may use multipolar
expansions \cite{multipol}.  However, for small separations one needs to
keep more and more terms in the expansion the smaller the distance; one
reaches a limit when the method is no longer feasible to use.  

In this work we study an approximation that comes handy in this situation.  The
Proximity Force Approximation (PFA). It was first used already in 1934 \cite {Derj} in connection with
coagulation of aerosols.  It is a very powerful and widely used
approximation for the interaction at short distances between two objects. 
Lamoreaux' experiment, discussed above, is interpreted within this
approximation.  It is difficult to make a strict estimate of how good the
approximation is but it has gained a wide-spread acceptance in recent
years.

The basic idea of the approximation is that the interaction potential
between the objects is an average interaction energy between parallel
planar interfaces,
%1
\begin{equation}
V\left( z \right) = \int_S {dSE_p \left( w \right)},
\label {PFA}
\end{equation}
where $E_p \left( w \right)$ is the interaction energy per unit area
between planar interfaces a distance $w$ apart.  The variable $z$ is the
closest distance between the two objects.  The surface $S$ is not uniquely
defined and the choice of $S$ is in the general case not a trivial choice to make.  In the
examples we discuss here the choice is more obvious.  For objects above a
substrate we may choose $S$ to be that part of a flat surface, parallel to
the substrate, that is covered by the projection of the object; $w$ is the
distance at $dS$ between the substrate and the object along the normal to
$S$.  In the case of two interacting objects we suggest that $S$ is that
part of the planar surface, perpendicular to the resulting force, where the
projections of the two objects overlap; $w$ is the distance at $dS$ between
the two objects along the normal to $S$. 

Deviations from the PFA results have been observed in experiments dealing
with spheres.\cite{Cap} There have been suggestions that the deviations are due 
to the thin coatings of the objects. However due to limitations in the experimental 
and theoretical parameters one can not say conclusively that 
the PFA calculations for coated plates give better predictions than PFA for 
solid and infinitely thick plates.

Some attempts have been made to show its validity in different cases. 
However, the lack of a general and rigorous proof still remains, as well as
estimates of the level of accuracy at all distances. In the present work we 
follow an alternative procedure where we perform an exhaustive study of 
the PFA results for a wide range of geometries and configurations, and 
compare them with the most exact results available in the literature.

 A general method for the calculation of van der Waals forces is shown in
 Ref.  \cite{RomSer}.  It is based on the solutions of a surface integral
 equation for the electric potential.  It is fast and robust, and allows
 for calculations with great accuracy. We use either this method or a multi-polar expansion method to compare with the PFA results. The multi-polar expansion method we use in the case of spheres and spheroids where the objects have simple symmetries. There are two versions of the integral equation method, one three-dimensional (3D) and one two-dimensional (2D). The 3D version is the more general of the two. In some situations one may use the 2D version which is faster converging. An example where this is possible is a finite cylinder above and parallel to a substrate. The force is in this case to a good approximation equal to the length of the cylinder times the force per unit length for an infinite cylinder above a substrate. In the infinite cylinder case we have cylinder symmetry and a 2D problem. The 3D and 2D versions give the same result for small separations but give different results for large. When they deviate it is the results from the 3D version that is correct.

We will in the sections that follow go through different examples and
compare the results to those from more elaborate calculations.  We use
extensions to the traditional PFA as discussed in Sec. \ref{PFAw}.  In Sec.
\ref{Planar} we show how the calculation of the interaction energy per unit
area in planar structures are performed. These results are used in our calculations for more general structures. In Sec. \ref{General} we give a 
general expression for the interaction energy for half spaces, cylinders and 
spheres. Sec. \ref{Sphere} is devoted to spherical objects,  Sec. 
\ref{Spheroids} to spheroids, Sec. \ref{Cylinder} to 
cylinders, Sec.  \ref{Cube} to cubes, Sec.  \ref{Cone} to cones, and Sec.  \ref{Wings} to wings. 
Finally, summary and conclusions are found in \ Sec.  \ref{Summary}. The
main analytical results are summarized in Table \ref{tab:table1} of Sec.
\ref{Summary}.

\section{\label{PFAw}The Proximity Force Approximation in this work}

\begin{figure}
\begin{minipage}{20pc}
\vspace{4.5pc}
\includegraphics[width=8cm]{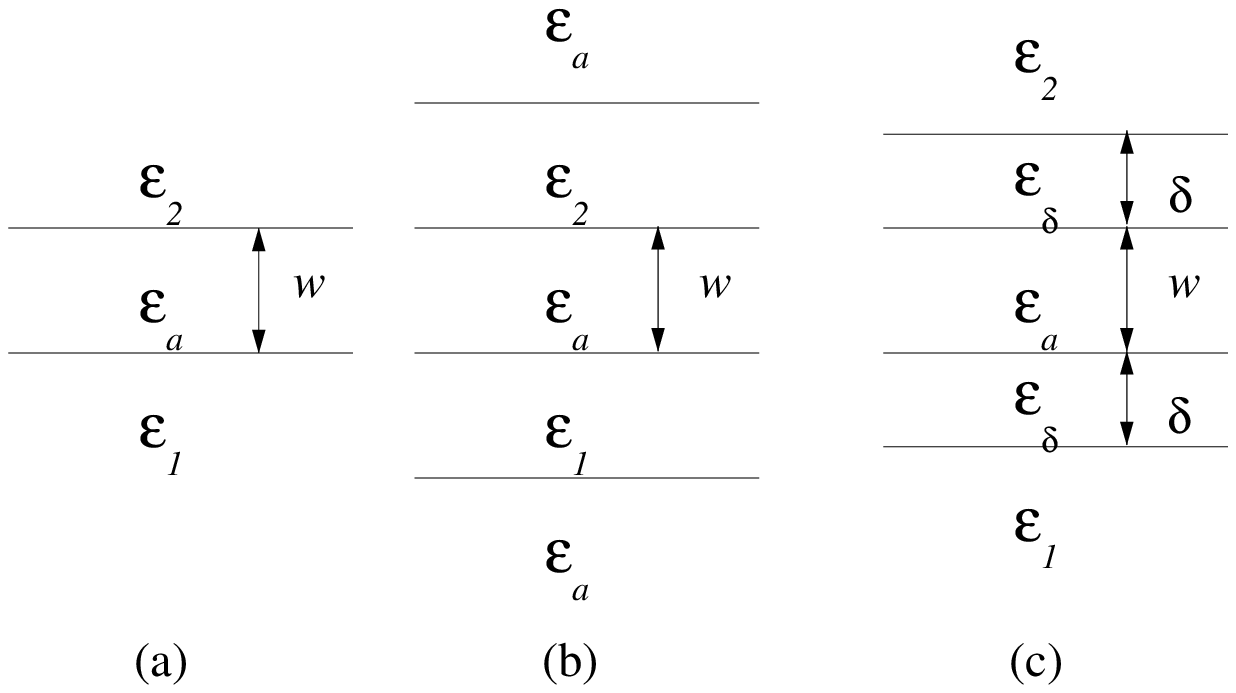}
\caption{(a) Planar structure entering $E_p \left( {w}\right)$ in the
traditional PFA. (b) Planar structure used in an extension of PFA to take
the back sides of the objects into account.  (c) Planar structure used in
the calculation of $E_p \left( {w,\delta }\right)$ in our treatment of
coated objects.}
\label{figu1}
\end{minipage}\hspace{1pc}
\begin{minipage}{20pc}
\includegraphics[width=8cm]{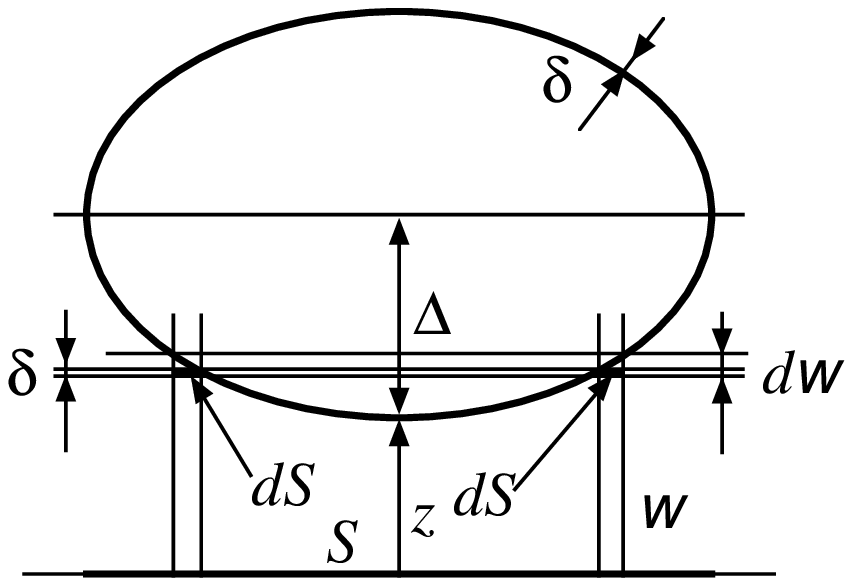}
\caption{Coated oblate spheroid above a substrate used to illustrate the 
parameters discussed in the text. }
\label{figu2}
\end{minipage}
\end{figure}
Traditionally in the PFA of the interaction between two objects one only
takes into consideration the surface of each object that is facing the
other object.  The function $E_p \left( {w}\right)$, in Eq. (\ref{PFA}), is
the energy per unit area for two half spaces, made up from the materials of
the two objects, separated by the distance $w$, see Fig. \ref{figu1}(a).  If
the objects are immersed in an ambient medium the gap between the two
half spaces is also filled with this medium.  With this treatment the
backsides of the objects have no effects at all.  One gets, e.g., the same
result for two spheres as for two halfspheres.  We know that the normal
modes contributing to the interaction may extend through the objects and
continue on the other side.  Thus, the backside may have important effects
on the results.  One may extend the treatment by using $E_p \left(
{w}\right)$ from a planar structure with four interfaces instead of two,
according to Fig. \ref{figu1}(b).  The distance between the interfaces
surrounding the object material is the local thickness of the object.  We
will not consider this in the present work, but address another limitation
that is more severe.  Two coated spheres, e.g., are in PFA treated as solid
spheres.  We will extend the treatment by considering a planar structure
with four interfaces, like in Fig. \ref{figu1}(c).  In the most general case
the two coatings are of different materials and of different thickness.  We
let $E_p \left( {w,\delta }\right)$ be the energy per unit area for two
coated half spaces, made up from the materials of the two objects, separated
the distance $w$.  The thickness of the coating is denoted by $\delta$.
We may express the potential as
%2
\begin{equation}
V\left( z \right) = \int_S {dSE_p \left( {w,\delta } \right)} =
\int\limits_z^{z + \Delta } {dw} \underbrace {\frac{{dS}}{{dw}}}_{g\left(
{w - z} \right)}E_p \left( {w,\delta } \right),
\end{equation}
and find the force as
%3
\begin{equation}
F\left( z \right) = - \frac{{dV}}{{dz}} = - g\left( \Delta \right)E_p
\left( {z + \Delta ,\delta } \right) + g\left( 0 \right)E_p \left(
{z,\delta } \right) - \int\limits_z^{z + \Delta } {dw} \frac{{dg\left( {w -
z} \right)}}{{dz}}E_p \left( {w,\delta } \right).
\label{PFAterms}
\end{equation}
In many cases $g\left( \Delta \right)$ vanishes, like in the illustrating
example in Fig. \ref{figu2}.  Here we have a coated spheroid above a
substrate.  The surface $S$ is defined through the projection of the object
onto the substrate.  When $g\left( \Delta \right)$ vanishes we have
%4
\begin{equation}
F\left( z \right) = g\left( 0 \right)E_p \left( {z,\delta } \right)\left[
{1 + \frac{1}{{g\left( 0 \right)E_p \left( {z,\delta }
\right)}}\int\limits_z^{z + \Delta } {dw} \frac{{d^2 S}}{{dw^2 }}E_p \left(
{w,\delta } \right)} \right],
\label{PFAFactors}
\end{equation}
where the first part is what one usually means with PFA. The remaining
part, within brackets, is a correction factor depending on the geometry. 
This factor is often dropped without any motivation at all or with the
argument that the resulting error is of the same order of magnitude as the
error in PFA itself.  Often $E_p \left( {w,\delta }\right)$ has a known
power dependence in $w$ over the whole integration interval.  Then we may
often find an analytical expression for the correction factor.  In Fig.
\ref{figu2} we study a coated oblate spheroid above an uncoated substrate. 
The lowest horizontal line is the substrate boundary.  The thicker part of
this line indicates the surface $S$, which is defined by the projection of
the spheroid on the substrate.  The integration variable $w$ varies from
$z$, the closest distance, to $z+\Delta$.  The region between $w$ and
$w+dw$ picks out a surface on the object whose projection on the substrate
is a ring of area $dS$.  This surface contributes, in the PFA spirit, to
the energy with $dS$ times the energy per unit area, $E_p \left( {w,\delta
}\right)$, of the planar configuration defined by the substrate and the
film of thickness $\delta$, indicated in the figure.  If the spheroid is
not an empty shell, the half space above the film should be filled by the
material of the spheroid.

To summarize, in this work we extend the traditional PFA in two ways; we
retain the correction factor of Eq.  (\ref{PFAFactors}); we extend PFA to
include the finite coat thickness of coated objects.  The effects of these
extensions are demonstrated and comparisons are made to numerical results
from more accurate calculations. Throughout the text we refer to results 
from traditional PFA as PFA results and to extended or corrected PFA 
results as full PFA results.

Before we proceed with the various objects we derive, in next section, the
interaction energy in those planar structures we need for the PFA
calculations.

\section{\label{Planar}Planar structures}
A general expression for the interaction energy per unit area in a planar
system where only one of the distances, $w$, between neighboring interfaces
is allowed to vary is \cite{Ser}
%5
\begin{equation}
\begin{array}{l}
 E_p \left( w \right) = \frac{\hbar }{2}\int\limits_{ - \infty }^\infty
 {\frac{{d\omega }}{{2\pi }}} \int {\frac{{d^2 k}}{{\left( {2\pi }
 \right)^2 }}} \left\{ {\ln \left[ {f\left( {{\bf{k}},i\omega ,w} \right)}
 \right] - \ln \left[ {f\left( {{\bf{k}},i\omega ,\infty } \right)}
 \right]} \right\} \\ \,\,\,\,\,\,\,\,\,\,\,\,\,\,\,\,\,\, = \frac{\hbar
 }{2}\int\limits_{ - \infty }^\infty {\frac{{d\omega }}{{2\pi }}} \int
 {\frac{{d^2 k}}{{\left( {2\pi } \right)^2 }}} \ln \left[ {\frac{{f\left(
 {{\bf{k}},i\omega ,w} \right)}}{{f\left( {{\bf{k}},i\omega ,\infty }
 \right)}}} \right], \end{array} 
\label{PlanarEn}
\end{equation}
where the reference energy is set to when $w$ is infinite.  The variable
{\bf k} is the two dimensional wave vector in the plane of the interfaces. 
The function ${f\left( {{\bf{k}},\omega ,w} \right)}$ is the function in
the condition for having electromagnetic normal modes in the planar system,
%6
\begin{equation}
f\left( {{\bf{k}},\omega ,w} \right) = 0.
\end{equation}
In the general case there are two of these functions, one for TE modes and
one for TM modes.  Here we limit the treatment to the non-retarded limit. 
One of the effects from neglecting retardation is that the TE modes are
absent.  In the case of a coated object above a coated substrate or of two
coated objects we need four interfaces between five regions.  For a
structure of the type $1|2|3|4|5$ where medium 3 has the variable thickness
$w$ we have
%7
\begin{equation}
\frac{{f\left( {{\bf{k}},\omega ,w} \right)}}{{f\left( {{\bf{k}},\omega
,\infty } \right)}} = 1 + e^{ - 2kw} \frac{{r_{23} r_{34} + e^{ - 2kd_2 }
r_{12} r_{34} + e^{ - 2kd_4 } r_{23} r_{45} + e^{ - 2kd_2 } e^{ - 2kd_4 }
r_{12} r_{45} }}{{1 + e^{ - 2kd_2 } r_{12} r_{23} + e^{ - 2kd_4 } r_{34}
r_{45} + e^{ - 2kd_2 } e^{ - 2kd_4 } r_{12} r_{23} r_{34} r_{45} }},
\label{FoverF}
\end{equation}
where $k = \left| {\bf{k}} \right|$ and
%8
\begin{equation}
r_{ij} = \frac{{\varepsilon _j \left( \omega \right) - \varepsilon _i
\left( \omega \right)}}{{\varepsilon _j \left( \omega \right) + \varepsilon
_i \left( \omega \right)}}.
\end{equation}
The function ${\varepsilon _i \left( \omega \right)}$ and $ d_i $ are the
dielectric function and thickness, respectively, of region $i$.  In the
case of a coated object above a substrate or of two objects where only one
is coated we need three interfaces between four regions.  For a structure of
the type $1|3|4|5$ where medium 3 has the variable thickness $w$ we let
${r_{12} }=0$ and get
%9
\begin{equation}
\frac{{f\left( {{\bf{k}},\omega ,w} \right)}}{{f\left( {{\bf{k}},\omega
,\infty } \right)}} = 1 + e^{ - 2kw} \frac{{r_{13} r_{34} + e^{ - 2kd_4 }
r_{13} r_{45} }}{{1 + e^{ - 2kd_4 } r_{34} r_{45} }}.
\end{equation}
In the case of an uncoated object above an uncoated substrate or of two
uncoated objects we need two interfaces between three regions.  For a
structure of the type $1|3|5$ where medium 3 has the variable thickness $w$
we let ${r_{45} }=0$ and get
%10
\begin{equation}
\frac{{f\left( {{\bf{k}},\omega ,w} \right)}}{{f\left( {{\bf{k}},\omega
,\infty } \right)}} = 1 + e^{ - 2kw} r_{13} r_{35} .
\end{equation}
For this last structure the integration over momentum in Eq. 
(\ref{PlanarEn}) may be performed and results in an infinite series.
%11
\begin{equation}
\begin{array}{l}
{\rm{E}}_{\rm{p}} \left( w \right) = \frac{\hbar }{{8\pi ^2 }}\int\limits_{
- \infty }^\infty {d\omega } \int\limits_0^\infty {dkk\ln \left[ {1 - e^{ -
2kw} r_{13} r_{53} } \right]} = \frac{\hbar }{{32\pi ^2 w^2 }}\int\limits_{
- \infty }^\infty {d\omega } \int\limits_0^\infty {dkk\ln \left[ {1 - e^{ -
k} r_{13} r_{53} } \right]} \\ \,\,\,\,\,\,\,\,\,\,\,\,\,\,\,\,\,\, = -
\frac{\hbar }{{32\pi ^2 w^2 }}\int\limits_{ - \infty }^\infty {d\omega }
\sum\limits_{l = 1}^\infty {\frac{{\left( {r_{13} r_{53} } \right)^l
}}{{l^3 }}} = - \frac{\hbar }{{32\pi ^2 w^2 }}\sum\limits_{l = 1}^\infty
{\frac{{\left\langle {\omega _l } \right\rangle }}{{l^3 }}},
\end{array}
\label{Series}
\end{equation}
where the characteristic dielectric integrals $\left\langle {\omega _l }
\right\rangle$ are
%12
\begin{equation}
\left\langle {\omega _l } \right\rangle = \int\limits_{ - \infty }^\infty
{d\omega \left[ {\frac{{\varepsilon _1 \left( \omega \right) - \varepsilon
_3 \left( \omega \right)}}{{\varepsilon _1 \left( \omega \right) +
\varepsilon _3 \left( \omega \right)}}\frac{{\varepsilon _5 \left( \omega
\right) - \varepsilon _3 \left( \omega \right)}}{{\varepsilon _5 \left(
\omega \right) + \varepsilon _3 \left( \omega \right)}}} \right]^l } .
\label{Characteristic}
\end{equation}

This result is obtained after a variable substitution, series expansion of
the logarithm, followed by the integration over momentum.
\section{\label{General}General expression for half spaces, cylinders and
spheres} For solid objects in neglect of the geometrical correction in Eq.
(\ref{PFAFactors}) one may find the following general expression for
half spaces, cylinders and spheres: \cite{Ser, Lang}
%13
\begin{equation}
E\left( z \right) = - \frac{\hbar }{{32\pi ^2 z^{\left( {1 + {n
\mathord{\left/ {\vphantom {n 2}} \right.  \kern-\nulldelimiterspace} 2}}
\right)} }}\Gamma \left( {1 + {n \mathord{\left/ {\vphantom {n 2}} \right. 
\kern-\nulldelimiterspace} 2}} \right)\left[ {\frac{{2\pi R_1 R_2 }}{{R_1+
R_2 }}} \right]^{1 - {n \mathord{\left/ {\vphantom {n 2}} \right. 
\kern-\nulldelimiterspace} 2}} \sum\limits_{l = 1}^\infty
{\frac{{\left\langle {\omega _l } \right\rangle }}{{l^3 }}} ,
\end{equation}
where $n=0$ for spheres, $n=1$ for cylinders, and $n=2$ for half spaces. 
The result is for the interaction energy in case of spheres, the
interaction energy per unit length for cylinders and the interaction energy
per unit area for half spaces.  The variable $z$ is the closest distance
between the objects.  Thus, for two half spaces we have
%14
\begin{equation}
E\left( z \right) = - \frac{\hbar }{{32\pi ^2 z^2 }}\sum\limits_{l =
1}^\infty {\frac{{\left\langle {\omega _l } \right\rangle }}{{l^3 }}} = E_p
\left( z \right),
\label{half spaces}
\end{equation}
for two cylinders of radii ${R_1 }$ and ${R_2 }$
%15
\begin{equation}
E\left( z \right) = - \frac{\hbar }{{32\pi ^2 z^{{3 \mathord{\left/
{\vphantom {3 2}} \right.  \kern-\nulldelimiterspace} 2}} }}\Gamma \left(
{{3 \mathord{\left/ {\vphantom {3 2}} \right.  \kern-\nulldelimiterspace}
2}} \right)\left[ {\frac{{2\pi R_1 R_2 }}{{R_1 + R_2 }}} \right]^{{1
\mathord{\left/ {\vphantom {1 2}} \right.  \kern-\nulldelimiterspace} 2}}
\sum\limits_{l = 1}^\infty {\frac{{\left\langle {\omega _l } \right\rangle
}}{{l^3 }}} = \Gamma \left( {{3 \mathord{\left/ {\vphantom {3 2}} \right. 
\kern-\nulldelimiterspace} 2}} \right)\sqrt {\frac{{2\pi R_1 R_2 z}}{{R_1 +
R_2 }}} E_p \left( z \right),
\label{cylinders}
\end{equation}
and for two spheres of radii ${R_1 }$ and ${R_2 }$ the result is
%16
\begin{equation}
E\left( z \right) = - \frac{\hbar }{{32\pi ^2 z}}\left[ {\frac{{2\pi R_1
R_2 }}{{R_1 + R_2 }}} \right]\sum\limits_{l = 1}^\infty
{\frac{{\left\langle {\omega _l } \right\rangle }}{{l^3 }}} = \left[
{\frac{{2\pi R_1 R_2 z}}{{R_1 + R_2 }}} \right]E_p \left( z \right).
\label{spheres}
\end{equation}
To get the result for a cylinder of radius $R$ above a substrate we let
${R_2 }$ go to infinity and replace ${R_1 }$ with $R$ in Eq.
(\ref{cylinders}).  This results in
%17
\begin{equation}
E\left( z \right) = - \frac{\hbar }{{32\pi ^2 z^{{3 \mathord{\left/
{\vphantom {3 2}} \right.  \kern-\nulldelimiterspace} 2}} }}\Gamma \left(
{{3 \mathord{\left/ {\vphantom {3 2}} \right.  \kern-\nulldelimiterspace}
2}} \right)\sqrt {2\pi R} \sum\limits_{l = 1}^\infty {\frac{{\left\langle
{\omega _l } \right\rangle }}{{l^3 }}} = \Gamma \left( {{3 \mathord{\left/
{\vphantom {3 2}} \right.  \kern-\nulldelimiterspace} 2}} \right)\sqrt
{2\pi Rz} E_p \left( z \right) .
\label{cylinderplane}
\end{equation}
\begin{figure}
\includegraphics[width=10cm]{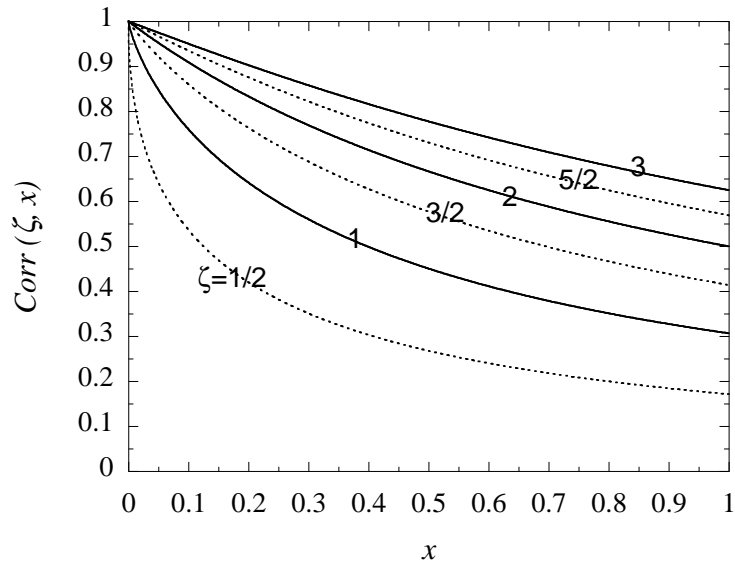}
\caption{The correction factor $Corr({\zeta}, x)$ for a set of 
${\zeta}$ values.}
\label{figu3}
\end{figure}
To get the result for a sphere of radius $R$ above a substrate we let ${R_2
}$ go to infinity and replace ${R_1 }$ with $R$ in Eq. (\ref{spheres}). 
This results in
%18
\begin{equation}
E\left( z \right) = - \frac{\hbar }{{32\pi ^2 z}}2\pi R\sum\limits_{l =
1}^\infty {\frac{{\left\langle {\omega _l } \right\rangle }}{{l^3 }}} =
2\pi RzE_p \left( z \right).
\end{equation}
This is as far as we get with the general formula.  Now, we continue in
next and the following sections with the geometrical corrections.  We start
with spherical objects.

\section{\label{Sphere}Spherical objects}

Spherical objects are often used in experiments. The advantage is that one 
avoids the problem of alignment. This is the case both for a sphere above 
a substrate and for two interacting spheres. In Lamoreaux's \cite{Lamo} classical 
measurement of the force between two gold plates one actually measured the 
force between a sphere and a planar surface.

\subsection{Sphere-substrate interaction}

\begin{figure}
\begin{minipage}{20pc}
\vspace{0.5pc}
\includegraphics[width=8cm]{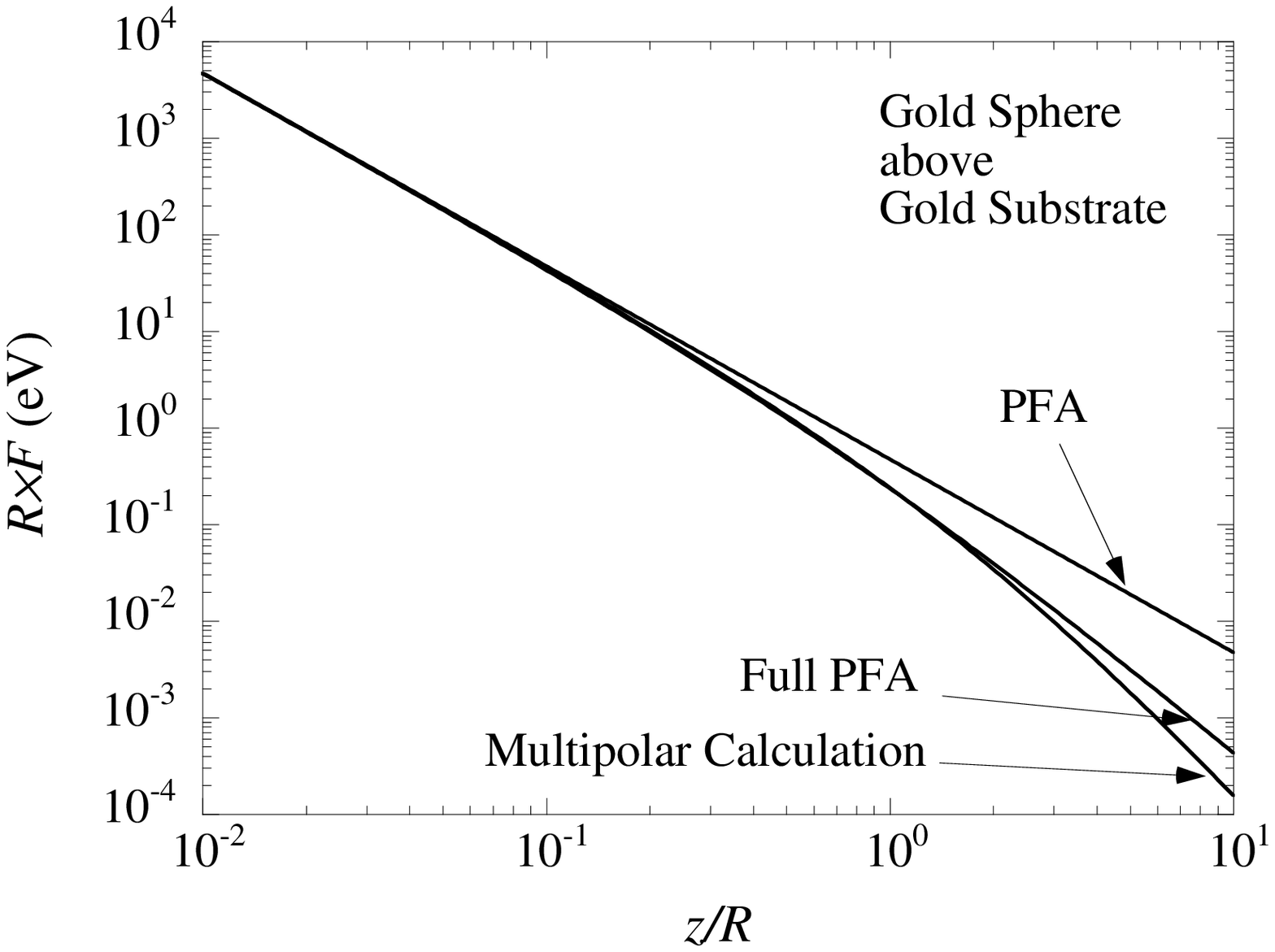}
\vspace{0.5pc}
\caption{$R$ times the non-retarded force on a gold sphere above a gold substrate
as function of $z/R$ for the multipolar result, the PFA result and the
geometry corrected or full PFA result.}
\label{figu4}
\end{minipage}\hspace{1pc}
\begin{minipage}{20pc}
\includegraphics[width=8cm]{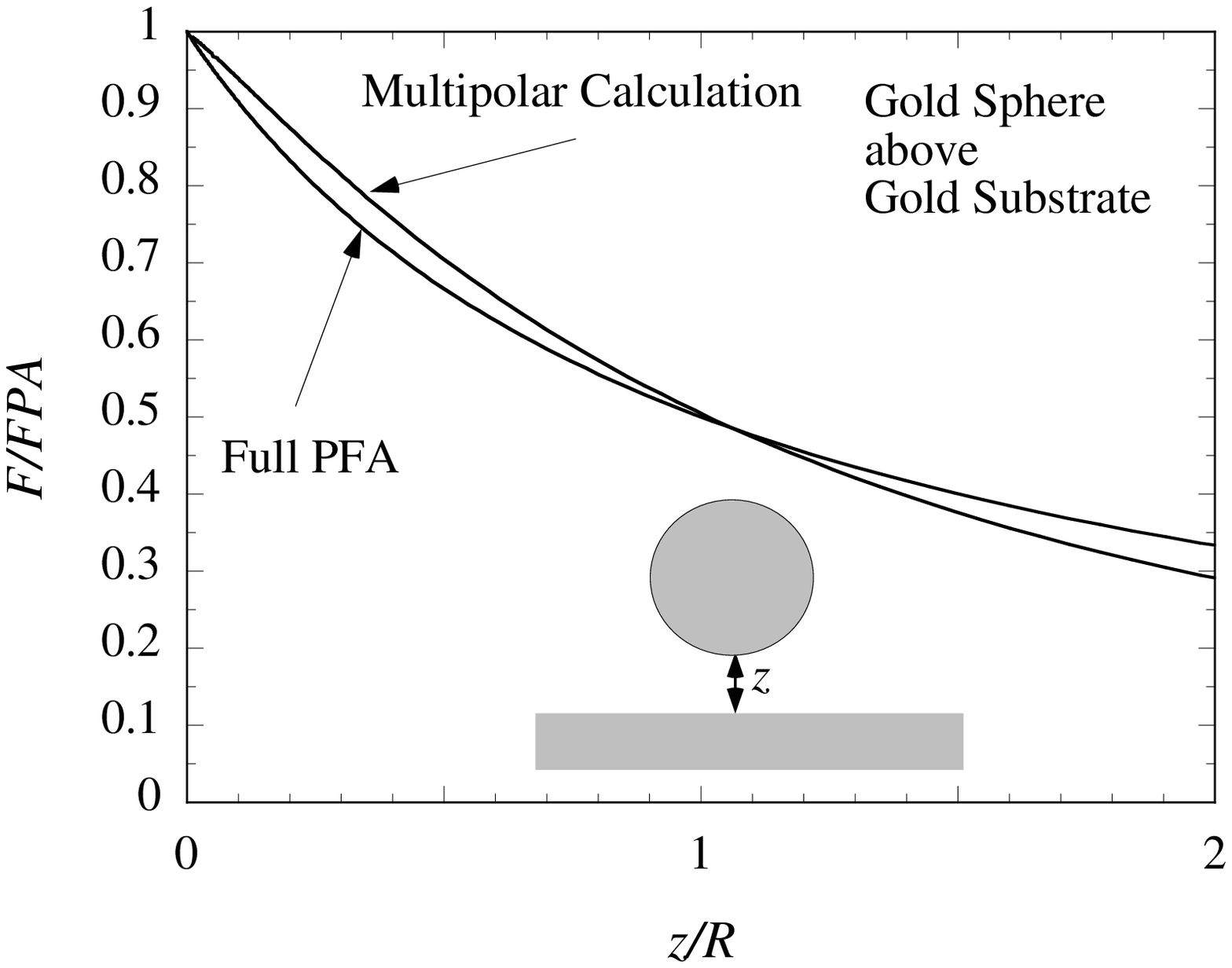}
\caption{The non-retarded force on a gold sphere above a gold substrate
relative the PFA result as function of $z/R$ for the
multipolar result and the full PFA result. }
\label{figu5}
\end{minipage}
\end{figure}

For a sphere above a substrate the parameters entering Eq.
(\ref{PFAFactors}) are
%19
\begin{equation}
\Delta = R;\quad g\left( x \right) = 2\pi \left( {R - x} \right);\quad
g\left( 0 \right) = 2\pi R;\quad \frac{{d^2 S}}{{dw^2 }} = - 2\pi,
\end{equation}
and this results in
%20
\begin{equation}
F\left( z \right) = 2\pi RE_p \left( {z,\delta } \right)\left[ {1 -
\frac{1}{{RE_p \left( {z,\delta } \right)}}\int\limits_z^{z + R} {dw} E_p
\left( {w,\delta } \right)} \right],
\label{FSphere}
\end{equation}
where we have included the possibility for the sphere and/or the substrate to have a coating of thickness $\delta$. If $E_p\left( {w,\delta } \right)$ varies as
%21
\begin{equation}
E_p \left( {w,\delta } \right) = {{ - C} \mathord{\left/ {\vphantom {{ - C}
{w^n }}} \right.  \kern-\nulldelimiterspace} {w^\zeta }}
\label{Evariation}
\end{equation}
in the whole integration interval we find the correction factor, $Corr 
({\zeta} ,x)$ on analytical form
%22
\begin{equation}
Corr\left( {\zeta ,x} \right) = \left\{ \begin{array}{l} 1 - {{\left[ {x -
{{x^\zeta } \mathord{\left/ {\vphantom {{x^\zeta } {\left( {1 + x}
\right)^{\zeta - 1} }}} \right.  \kern-\nulldelimiterspace} {\left( {1 + x}
\right)^{\zeta - 1} }}} \right]} \mathord{\left/ {\vphantom {{\left[ {x -
{{x^\zeta } \mathord{\left/ {\vphantom {{x^\zeta } {\left( {1 + x}
\right)^{\zeta - 1} }}} \right.  \kern-\nulldelimiterspace} {\left( {1 + x}
\right)^{\zeta - 1} }}} \right]} {\left( {\zeta - 1} \right)}}} \right. 
\kern-\nulldelimiterspace} {\left( {\zeta - 1} \right)}};\quad \zeta \ne 1
\\ 1 - x\ln \left[ {{{\left( {1 + x} \right)} \mathord{\left/ {\vphantom
{{\left( {1 + x} \right)} x}} \right.  \kern-\nulldelimiterspace} x}}
\right];\quad \zeta = 1 \\ \end{array} \right.
\label{Corr}
\end{equation}
where $x = {z \mathord{\left/{\vphantom {z R}} \right. 
\kern-\nulldelimiterspace} R}$. The correction factor is illustrated for 
some ${\zeta}$ values in Fig. \ref{figu3}.
Eq.  (\ref{FSphere}) is valid both in the non-retarded and retarded
separation regions.  The PFA is only good for small $x$-values but if $R$
is big enough one may still be in the retarded region.  For uncoated sphere
and substrate $\zeta =2$ ($\zeta =3$) in the non-retarded (retarded)
region.  If either of the sphere and substrate or both are coated the
separation dependence is in general more complex.  If a metallic coating is
thin enough there is a separation range where $\zeta =5/2$.  

       With the proper scaling one may produce universal figures, i.e., figures
that are identical for all values of $R$.  From Eqs.  ({\ref{PlanarEn}})
and (\ref{FoverF}) follows that $ E_p \left( {w,\delta } \right) =
{{f\left( {{w \mathord{\left/ {\vphantom {w R}} \right.
\kern-\nulldelimiterspace} R},{\delta \mathord{\left/ {\vphantom {\delta
R}} \right.  \kern-\nulldelimiterspace} R}} \right)} \mathord{\left/
{\vphantom {{f\left( {{w \mathord{\left/ {\vphantom {w R}} \right. 
\kern-\nulldelimiterspace} R},{\delta \mathord{\left/ {\vphantom {\delta
R}} \right.  \kern-\nulldelimiterspace} R}} \right)} {w^2 }}} \right. 
\kern-\nulldelimiterspace} {w^2 }}$. From Eq. (\ref{FSphere}) follows that 
if one plots $R \times F$ as a function of $z/R$ the results look identical
as long as the coat thickness is the same fraction of $R$.  This means that
there is no need to repeat the calculations for different sphere radii.  In
Fig.  \ref{figu4} we show $R$ times the non-retarded force on a gold sphere
above a gold substrate as function of normalized separation $z/R$. 
Throughout we use the dielectric function of gold given in Ref. 
\cite{goldlamb}.
The differences between the curves are not clearly seen in a figure like 
this. In Fig.  \ref{figu5} we show the result from the multipolar 
calculation and the geometry corrected or full PFA result relative the PFA result. We see that  the full PFA means a substantial improvement from PFA.
\begin{figure}
\includegraphics[width=10cm]{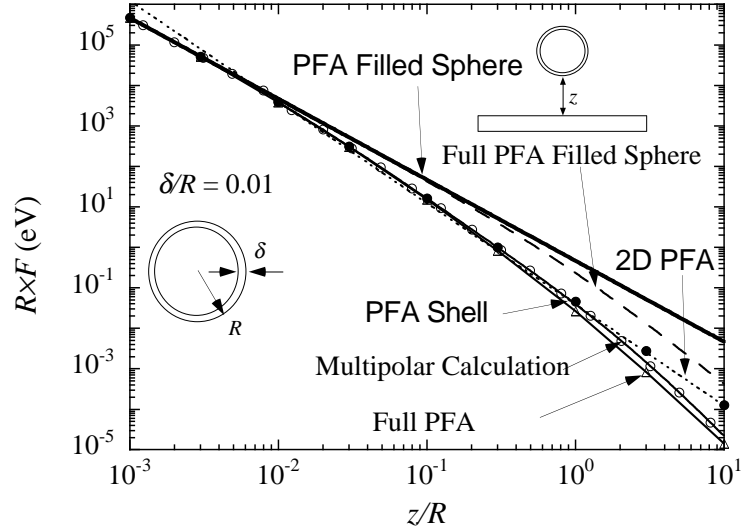}
\caption{$R$ times the non-retarded force on a spherical gold shell above a
gold substrate as function of $z/R$.  The curves are for the multipolar
result (thin solid curve with open circles), the PFA result for a solid
sphere (thick solid curve), the full PFA result for a solid
sphere (dashed curve), the two-dimensional PFA result (dotted curve),
the PFA result for the shell (solid circles) and the full PFA 
for the spherical shell (triangles).}
\label{figu6}
\end{figure}
In Fig.  \ref{figu6} we show $R$ times the non-retarded force on a
spherical gold shell above a gold substrate as function of normalized
separation $z/R$.  The thick solid curve and dashed curve are the PFA and
full PFA curves, respectively, for a solid gold sphere and are the same as
in Fig.  \ref{figu4}; the solid curve with open circles is the result from
the multipolar calculation for a shell of a thickness of one percent of the
radius; the filled circles are our extended PFA result for the gold shell;
the solid curve with triangles is our extended and full PFA result for the
gold shell.  We note that the full result, from the multipolar calculation,
follows the PFA result for a solid gold sphere for distances smaller than
approximately the coat thickness.  Then for larger separations there is a
region where it follows rather closely the 2D PFA result obtained by
replacing the gold film with a 2D (two dimensional) metallic sheet with 2D
electron density given as the projection of the 3D electron density of the
film.  These results are derived in Refs.  \cite{BoSer,SerBjo},
%23
\begin{equation}
 R \times F = 2\pi R^2 E_p \left( z \right) \approx 0.1556\sqrt {{{n\hbar
 ^2 e^2 } \mathord{\left/ {\vphantom {{n\hbar ^2 e^2 } {m_e }}} \right. 
 \kern-\nulldelimiterspace} {m_e }}} {{\sqrt {{\delta \mathord{\left/
 {\vphantom {\delta R}} \right.  \kern-\nulldelimiterspace} R}} }
 \mathord{\left/ {\vphantom {{\sqrt {{\delta \mathord{\left/ {\vphantom
 {\delta R}} \right.  \kern-\nulldelimiterspace} R}} } {\left( {{z
 \mathord{\left/ {\vphantom {z R}} \right.  \kern-\nulldelimiterspace} R}}
 \right)^{{5 \mathord{\left/ {\vphantom {5 2}} \right. 
 \kern-\nulldelimiterspace} 2}} }}} \right.  \kern-\nulldelimiterspace}
 {\left( {{z \mathord{\left/ {\vphantom {z R}} \right. 
 \kern-\nulldelimiterspace} R}} \right)^{{5 \mathord{\left/ {\vphantom {5
 2}} \right.  \kern-\nulldelimiterspace} 2}} }},
\end{equation}
where $n$ is the conduction electron density of gold.
\subsection{Sphere-sphere interaction}
For two spheres of equal size the parameters entering Eq.
(\ref{PFAFactors}) are
%24
\begin{equation}
\Delta = 2R;\quad g\left( x \right) = \pi \left( {R - x/2} \right);\quad
g\left( 0 \right) = \pi R;\quad \frac{{d^2 S}}{{dw^2 }} = -\pi /2,
\end{equation}
and this results in
%25
\begin{equation}
F\left( z \right) = \pi RE_p \left( {z,\delta } \right)\left[ {1 -
\frac{1}{{2RE_p \left( {z,\delta } \right)}}\int\limits_z^{z + 2R} {dw} E_p
\left( {w,\delta } \right)} \right].
\label{F2Sphere}
\end{equation}
\begin{figure}
\begin{minipage}{20pc}
\includegraphics[width=8cm]{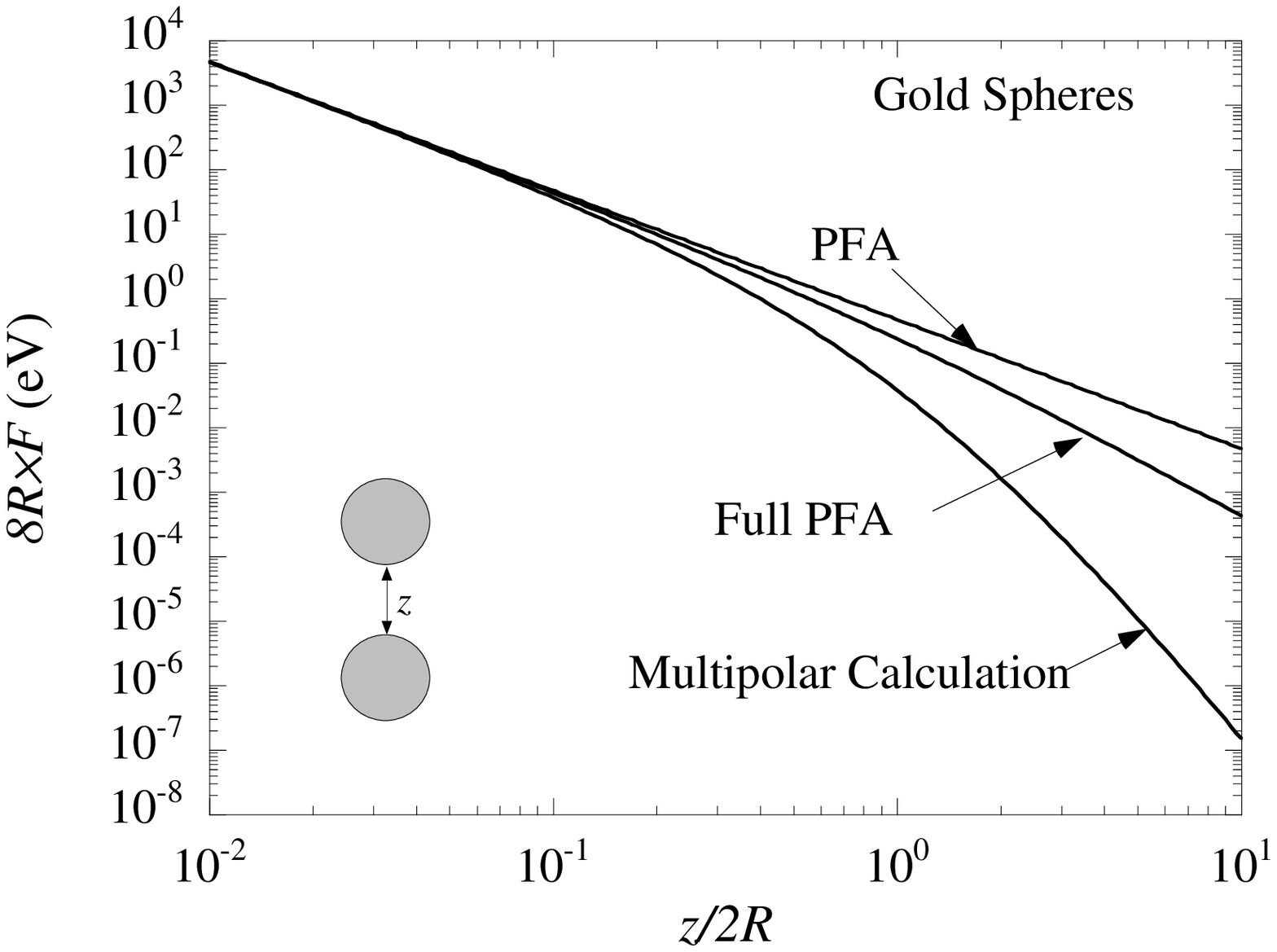}
\vspace{0.5pc}
\caption{$8R$ times the non-retarded force between two gold spheres as function of $z/2R$ for the multipolar result, the PFA result and the
full PFA result.}
\label{figu7}
\end{minipage}\hspace{1pc}
\begin{minipage}{20pc}
\includegraphics[width=8cm]{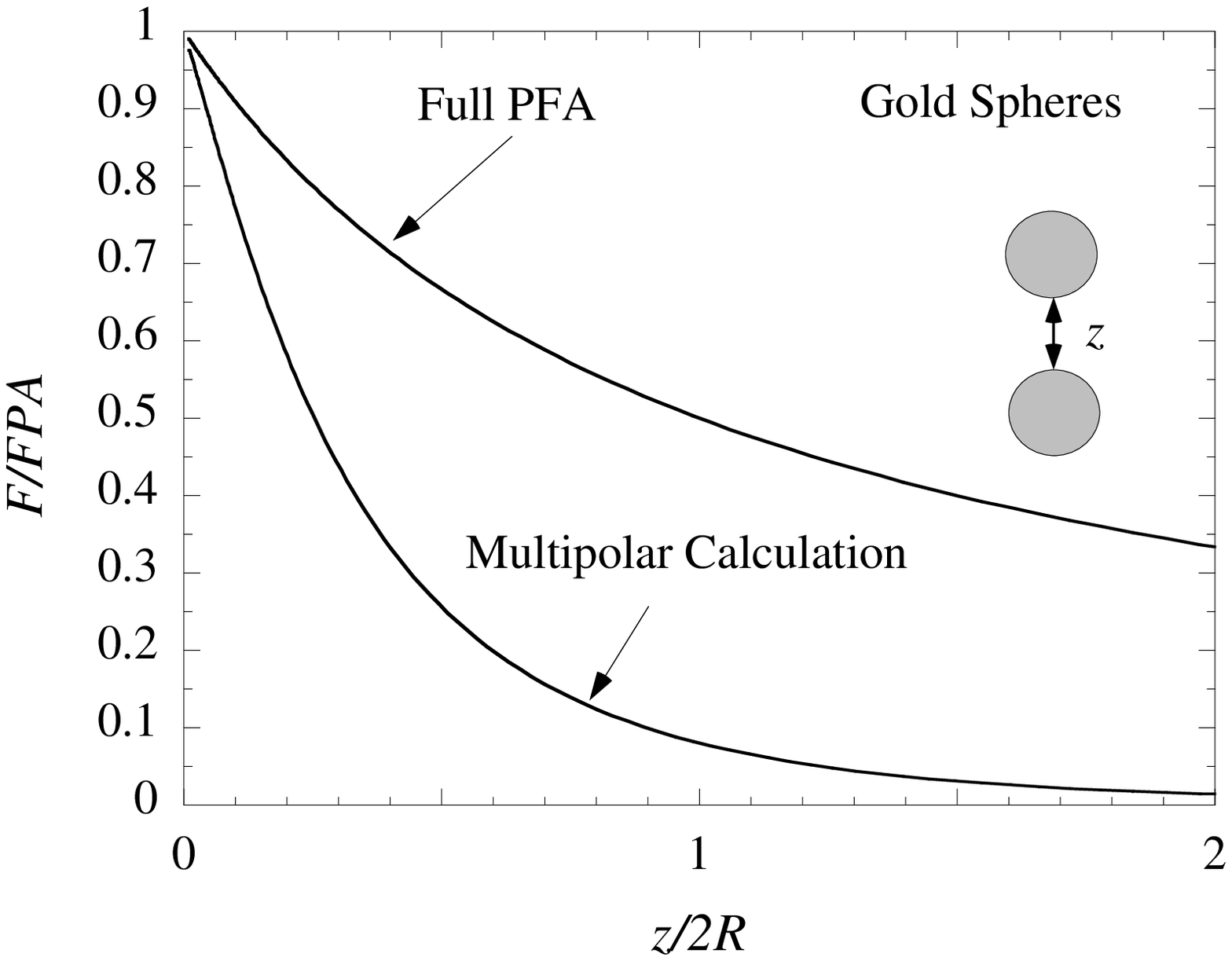}
\caption{The non-retarded force between two equal size gold spheres
relative the PFA result as function of $z/2R$ for the
multipolar result and the full PFA result. }
\label{figu8}
\end{minipage}
\end{figure}
If $E_p\left( {w,\delta } \right)$ varies as in Eq.  (\ref{Evariation}) in
the whole integration interval we again find the correction factor, $Corr
({\zeta} ,x)$, on analytical form and it is identical to the expression in
Eq.  (\ref{Corr}) but now $x = {z \mathord{\left/{\vphantom {z {2R}}}
\right.\kern-\nulldelimiterspace} {2R}}$.  In Fig.  \ref{figu7} we show $8R$
times the non-retarded force between two gold spheres of equal size as
function of normalized separation $z/2R$. Plotted in this way the PFA and 
full PFA curves are identical to the corresponding ones in Fig. 
\ref{figu4}; the result from the multipolar calculation is not.
In Fig.  \ref{figu8} we show the result from the multipolar 
calculation and the full PFA result relative the PFA result. We find that the agreement is not as good as in the case of a sphere above a substrate.
\section{\label{Spheroids}Spheroids}
A spheroid has an axis of symmetry. We let this axis be perpendicular to 
the substrate in case of a spheroid above a substrate; in the case of two 
spheroids we let both symmetry axes coincide with the line joining the 
centers of the spheroids. We let $y$ be the variable along the symmetry axis and $r$ 
the variable perpendicular to the symmetry axis. Let $A$ ($B$) be the 
larger (smaller) of the two semiaxes and let 
%26
\begin{equation}
\gamma  = {A \mathord{\left/
 {\vphantom {A B}} \right.
 \kern-\nulldelimiterspace} B} > 1.
\end{equation}
There are two types of spheroid, oblate defined by the equation
%27
\begin{equation}
\frac{{r^2 }}{{A^2 }} + \frac{{y^2 }}{{B^2 }} = 1,
\end{equation}
and prolate
defined by
%28
\begin{equation}
\frac{{r^2 }}{{B^2 }} + \frac{{y^2 }}{{A^2 }} = 1.
\end{equation}
The oblate spheroid varies between a sphere of radius $A$ when $B$ is equal
to $A$ and a disk of radius $A$ when $B$ is equal to $0$; the prolate
spheroid varies between a sphere of radius $A$ when $B$ is equal to $A$ and
a pin of length $A$ when $B$ is equal to $0$.

\subsection{Spheroid-substrate interaction}
We start with the oblate spheroids.  For an oblate spheroid above a
substrate the parameters entering Eq. (\ref{PFAFactors}) are
%29
\begin{equation}
\Delta = B;\quad g\left( x \right) = - 2\pi \gamma ^2 \left( {x - B}
\right);\quad g\left( 0 \right) = 2\pi \gamma ^2 B;\quad \frac{{d^2
S}}{{dw^2 }} = - 2\pi \gamma ^2,
\end{equation}
and this results in
%30
\begin{equation}
F\left( z \right) = 2\pi B\gamma ^2 E_p \left( {z,\delta } \right)\left[ {1
- \frac{1}{{BE_p \left( {z,\delta } \right)}}\int\limits_z^{z + B} {dw} E_p
\left( {w,\delta } \right)} \right].
\end{equation}
Thus,
%31
\begin{equation}
 F\left( z \right) = 2\pi B\gamma ^2 E_p \left( {z,\delta }
 \right)Corr\left( {\zeta ,x} \right);\quad x = {z \mathord{\left/
 {\vphantom {z B}} \right.  \kern-\nulldelimiterspace} B},
\end{equation}
if $ E_p \left( {z,\delta } \right)$ has a power law dependence. The 
correction factor, $Corr\left( {\zeta ,x} \right)$, is the same as in Eq. 
(\ref{Corr}). 

To get the results for prolate spheroids we just replace $B$ with $A$ and $
\gamma$ with $\gamma ^{ - 1}$. The result is
%32
\begin{equation}
F\left( z \right) = 2\pi A\gamma ^{-2} E_p \left( {z,\delta } \right)\left[ {1
- \frac{1}{{AE_p \left( {z,\delta } \right)}}\int\limits_z^{z + A} {dw} E_p
\left( {w,\delta } \right)} \right],
\end{equation}
and if $ E_p \left( {z,\delta } \right)$ has a power law dependence it is
%33
\begin{equation}
 F\left( z \right) = 2\pi A\gamma ^{-2} E_p \left( {z,\delta }
 \right)Corr\left( {\zeta ,x} \right);\quad x = {z \mathord{\left/
 {\vphantom {z A}} \right.  \kern-\nulldelimiterspace} A}.
\end{equation}
The correction factor, $Corr\left( {\zeta ,x} \right)$, is the same as in
Eq.  (\ref{Corr}).

\begin{figure}
\begin{minipage}{20pc}
\vspace{1.1pc}
\includegraphics[width=8.1cm]{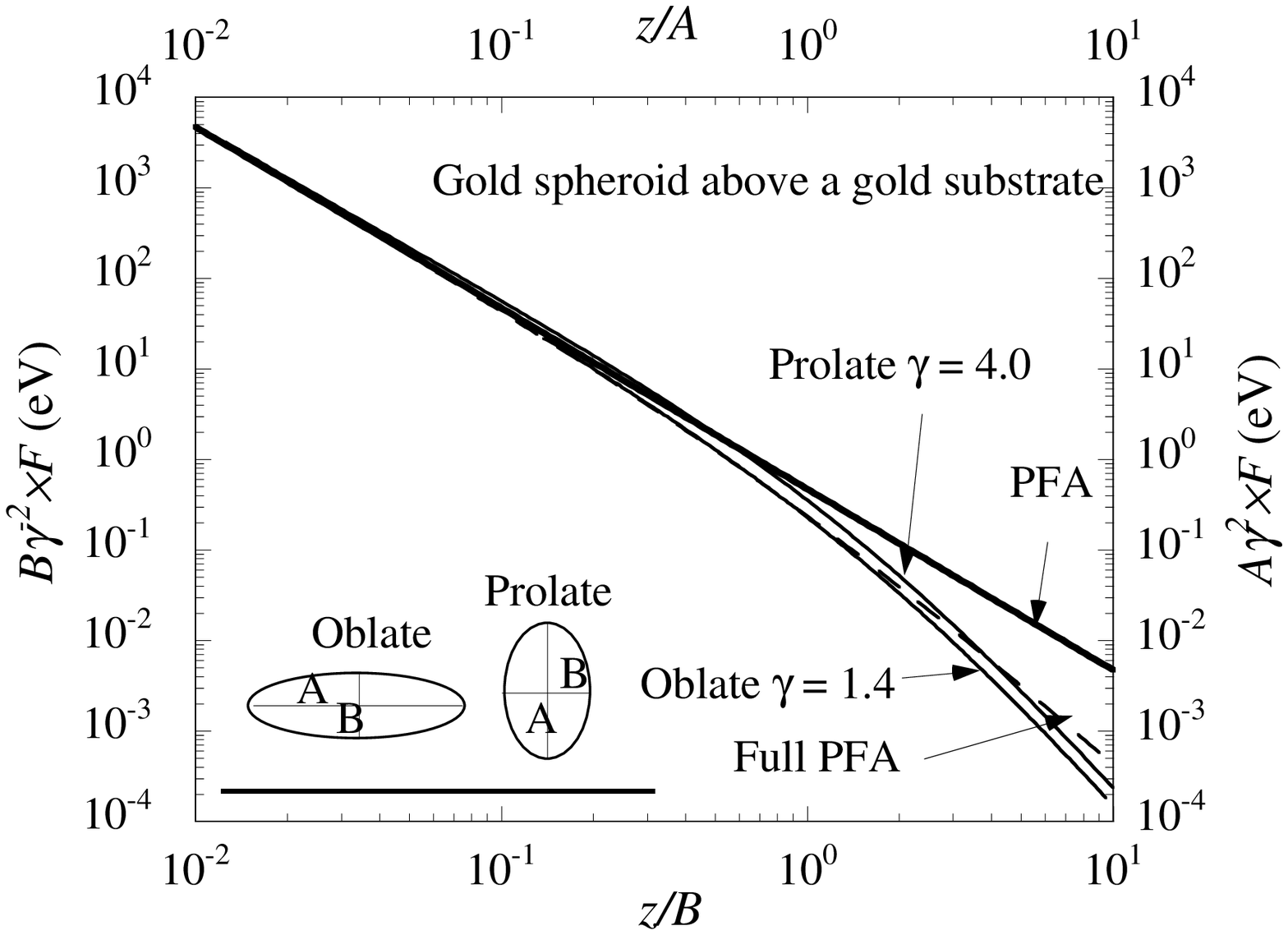}
\caption{The force on a gold spheroid above a gold substrate. The thick 
solid curve and dashed curve are for the PFA and full PFA,
respectively.  The upper (lower) solid curve is for a prolate (oblate)
spheroid with $\gamma = 4.0$ ($\gamma = 1.4$).  The left (right) and 
lower (upper) axes are for oblate (prolate) spheroids.}
\label{figu9}
\end{minipage}\hspace{1pc}
\begin{minipage}{20pc}
\includegraphics[width=8.1cm]{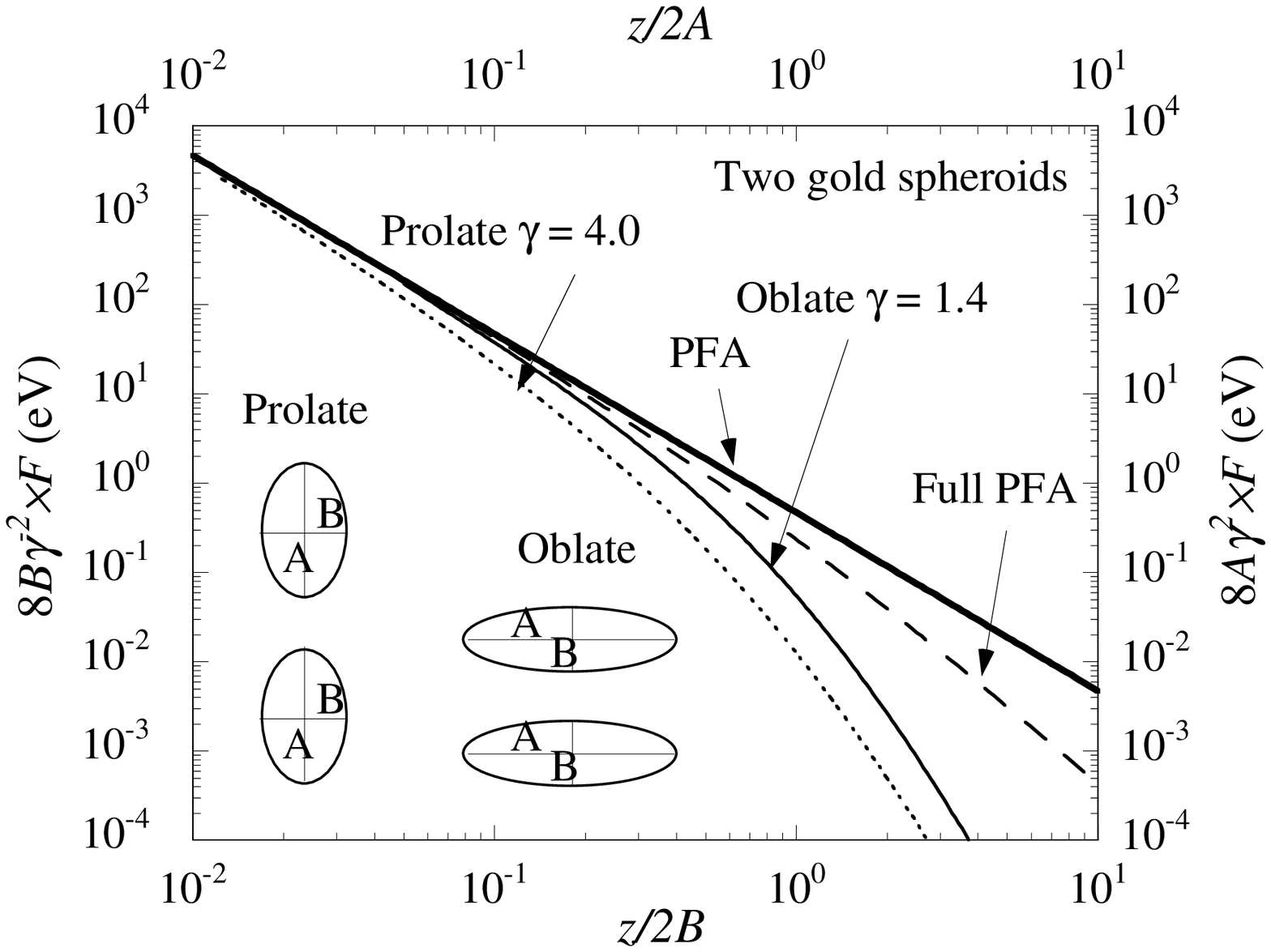}
\caption{The force between two gold spheroids. The thick 
solid curve and dashed curve are for the PFA and full PFA,
respectively.  The upper (lower) solid curve is for a prolate (oblate)
spheroid with $\gamma = 4.0$ ($\gamma = 1.4$).  The left (right) and 
lower (upper) axes are for oblate (prolate) spheroids.}
\label{figu10}
\end{minipage}
\end{figure}

 If one for an oblate (prolate) spheroid plots $B\gamma ^{ - 2} \times F$
 ($A\gamma ^{ 2} \times F$) as a function of $z/B$ ($z/A$) and lets the
 thickness of the coating be in units of $B$ ($A$) one obtains the same
 universal PFA and full PFA curves as we previously found for a sphere
 above a substrate or for two spheres.  Fig.  \ref{figu9} shows the results
 for an oblate (prolate) spheroid with $\gamma = 1.4$ ($\gamma = 4.0$)
 above a gold substrate.  An alternative way to plot the curves is to
 plot $\tilde R \times F$ as function of $ {z \mathord{\left/ {\vphantom
 {z {\tilde R}}} \right.\kern-\nulldelimiterspace} {\tilde R}}$, where $\tilde 
 R$ is the radius of curvature at the point of closest contact. Then the 
 PFA result is universal but the full PFA result is different for the 
 oblate and prolate cases and depends on the ${\gamma}$ value. The radius 
 of curvature is $B\gamma ^2$ ($A\gamma ^{ - 2} $) for oblate 
 (prolate) spheroids.
\subsection{Spheroid-spheroid interaction}
In the case two equal spheroids aligned along the common symmetry axis we 
find in the oblate case
%34
\begin{equation}
F\left( z \right) = \pi B\gamma ^2 E_p \left( {z,\delta } \right)\left[ {1
- \frac{1}{{2BE_p \left( {z,\delta } \right)}}\int\limits_z^{z + 2B} {dw} E_p
\left( {w,\delta } \right)} \right],
\end{equation}
and if $ E_p \left( {z,\delta } \right)$ has a power law dependence it is
%35
\begin{equation}
 F\left( z \right) = \pi B\gamma ^2 E_p \left( {z,\delta }
 \right)Corr\left( {\zeta ,x} \right);\quad x = {z \mathord{\left/
 {\vphantom {z 2B}} \right.  \kern-\nulldelimiterspace} 2B}.
\end{equation}
The correction factor, $Corr\left( {\zeta ,x} \right)$, is the same as in
Eq.  (\ref{Corr}).

For two prolate spheroids we obtain
%36
\begin{equation}
F\left( z \right) = \pi A\gamma ^{-2} E_p \left( {z,\delta } \right)\left[ {1
- \frac{1}{{2AE_p \left( {z,\delta } \right)}}\int\limits_z^{z + 2A} {dw} E_p
\left( {w,\delta } \right)} \right],
\end{equation}
and
%37
\begin{equation}
 F\left( z \right) = \pi A\gamma ^{-2} E_p \left( {z,\delta }
 \right)Corr\left( {\zeta ,x} \right);\quad x = {z \mathord{\left/
 {\vphantom {z 2A}} \right.  \kern-\nulldelimiterspace} 2A},
\end{equation}
 if $ E_p \left( {z,\delta } \right)$ has a power law dependence.  The
 correction factor, $Corr\left( {\zeta ,x} \right)$, is the same as in Eq.
 (\ref{Corr}).  If one for two oblate (prolate) spheroids plots $8B\gamma
 ^{ - 2} \times F$ ($8A\gamma ^{ 2} \times F$) as a function of $z/2B$
 ($z/2A$) and lets the thickness of the coating be in units of $2B$ ($2A$)
 one obtains the same universal PFA and full PFA curves as we
 previously found for a sphere above a substrate, for two spheres and for a
 spheroid above a substrate. In Fig. \ref{figu10} we show the force 
 between two gold spheroids, two prolate with $\gamma=4.0$ and two oblate 
 with $\gamma=1.4$. An alternative way to plot the curves is to plot $8\tilde
 R \times F$ as function of $ {z \mathord{\left/ {\vphantom {z {\tilde 2R}}}
 \right.\kern-\nulldelimiterspace} {\tilde 2R}}$, where $\tilde R$ is the
 radius of curvature at the point of closest contact.  Then the PFA result
 is universal but the full PFA result is different for the oblate and
 prolate cases and depends on the ${\gamma}$ value. 

\section{\label{Cylinder} Cylinders}

The problem of alignment in measurements of the dispersion forces is partly avoided by using cylinders. It is fully 
avoided in case of crossed cylinders \cite{Ede}

\subsection{Cylinder-substrate interaction}
\begin{figure}
\includegraphics[width=10cm]{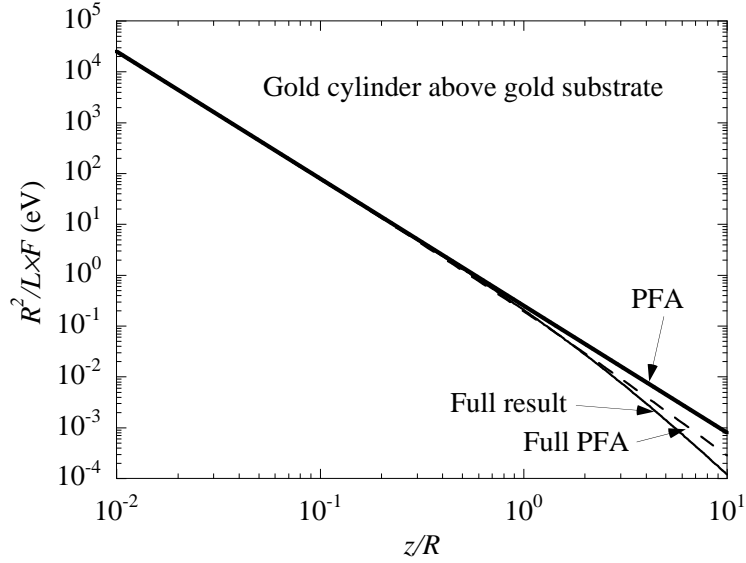}
\caption{The force on a gold cylinder of radius $R$ and length $L$ above a
gold substrate.  The thick solid curve and dashed curve are for the PFA and
full PFA, respectively.  The thin solid curve is the full result from
calculations based on integral equation techniques.}
\label{figu11}
\end{figure}

Here, we consider a cylinder of radius $R$ and length $L$ above a 
substrate. The closest distance is $z$, as before. This is a geometry where 
once again $g\left( \Delta \right)$ of Eq. (\ref{PFAterms}) vanishes. 
However, here one runs into another problem, viz., that $g\left( 0 \right)$ 
diverges. We need to make an alternative derivation,
%38
\begin{equation}
\begin{array}{l}
 V\left( z \right) = \int\limits_S {dS} E_p \left( w \right) =
 2L\int\limits_0^R {dy} E_p \left( {z + R - \sqrt {R^2 - y^2 } } \right) =
 \left| {w = z + R - \sqrt {R^2 - y^2 } } \right| \\
 \,\,\,\,\,\,\,\,\,\,\,\,\,\, = 2L\int\limits_z^{z + R} {dw} \frac{{z + R -
 w}}{{\sqrt {R^2 - \left( {z + R - w} \right)^2 } }}E_p \left( w \right) =
 \left| {w \to u + z} \right| = 2L\int\limits_0^R {du} \frac{{R -
 u}}{{\sqrt {R^2 - \left( {R - u} \right)^2 } }}E_p \left( {u + z} \right),
 \end{array}
\end{equation}
and
%39
\begin{equation}
F\left( z \right) = - \frac{{dV\left( z \right)}}{{dz}} = -
2L\int\limits_0^R {du} \frac{{R - u}}{{\sqrt {R^2 - \left( {R - u}
\right)^2 } }}\frac{{dE_p \left( {u + z} \right)}}{{dz}}.
\end{equation}
We limit the treatment to power law dependence in the non-retarded limit 
and obtain
%40
\begin{equation}
\begin{array}{l}
 F\left( z \right) = \left[ {LE_p \left( z \right)\left( {{{3\sqrt \pi }
 \mathord{\left/ {\vphantom {{3\sqrt \pi } 4}} \right. 
 \kern-\nulldelimiterspace} 4}} \right)\sqrt {{{2\pi R} \mathord{\left/
 {\vphantom {{2\pi R} z}} \right.  \kern-\nulldelimiterspace} z}} } \right]
 \\ \,\,\,\,\,\,\,\,\,\,\,\,\,\,\,\,\,\, \times \left[ {\frac{{6\left( {x +
 1} \right)^2 \tan ^{ - 1} \left( {{1 \mathord{\left/ {\vphantom {1 {\sqrt
 {x^2 + 2x} }}} \right.  \kern-\nulldelimiterspace} {\sqrt {x^2 + 2x} }}}
 \right) + 3\pi \left( {x + 1} \right)^2 + 2\sqrt {x^2 + 2x} \left( {2x^2 +
 4x + 3} \right)}}{{\left( {{{3\sqrt \pi } \mathord{\left/ {\vphantom
 {{3\sqrt \pi } 4}} \right.  \kern-\nulldelimiterspace} 4}} \right)\sqrt
 {{{2\pi } \mathord{\left/ {\vphantom {{2\pi } x}} \right. 
 \kern-\nulldelimiterspace} x}} \left( {x + 1} \right)\left( {x + 2}
 \right)^2 \sqrt {x^2 + 2x} }}} \right],
\end{array}
\end{equation}
or
%41
\begin{equation}
\begin{array}{l}
 {{F\left( z \right)R^2 } \mathord{\left/ {\vphantom {{F\left( z \right)R^2
 } L}} \right.  \kern-\nulldelimiterspace} L} = \left[ {E_p \left( x
 \right)\left( {{{3\sqrt \pi } \mathord{\left/ {\vphantom {{3\sqrt \pi }
 4}} \right.  \kern-\nulldelimiterspace} 4}} \right)\sqrt {{{2\pi }
 \mathord{\left/ {\vphantom {{2\pi } x}} \right. 
 \kern-\nulldelimiterspace} x}} } \right] \\
 \,\,\,\,\,\,\,\,\,\,\,\,\,\,\,\,\,\, \times \left[ {\frac{{6\left( {x + 1}
 \right)^2 \tan ^{ - 1} \left( {{1 \mathord{\left/ {\vphantom {1 {\sqrt
 {x^2 + 2x} }}} \right.  \kern-\nulldelimiterspace} {\sqrt {x^2 + 2x} }}}
 \right) + 3\pi \left( {x + 1} \right)^2 + 2\sqrt {x^2 + 2x} \left( {2x^2 +
 4x + 3} \right)}}{{\left( {{{3\sqrt \pi } \mathord{\left/ {\vphantom
 {{3\sqrt \pi } 4}} \right.  \kern-\nulldelimiterspace} 4}} \right)\sqrt
 {{{2\pi } \mathord{\left/ {\vphantom {{2\pi } x}} \right. 
 \kern-\nulldelimiterspace} x}} \left( {x + 1} \right)\left( {x + 2}
 \right)^2 \sqrt {x^2 + 2x} }}} \right],
\label{CylSub}
\end{array}
\end{equation}
where we have written the results on universal form.  The first factor is
the PFA result and the second the correction factor from the full geometry. 
The variable $x=z/R$. The result is valid under the assumption that $E_p
\left( z \right) \propto z^{ - 2}$.  Fig.  \ref{figu11} shows the result
(thin solid curve) for a gold cylinder above a gold substrate as calculated
with integral equation techniques \cite{RomSer}.  We have performed
calculations for an infinite cylinder and for finite ones with $L=R$ and
$L=2R$.  All three curves fall within the curve thickness on a plot like
this.  The result from the infinite cylinder calculation we limited to $z<R$ otherwise we would have gotten deviations. The thick solid (thin dashed) curve is the PFA (full PFA)
result.
\subsection{Cylinder-cylinder interaction}
The result for the force between two identical parallel cylinders is easily 
obtained from the results in the previous subsection. It is
%42
\begin{equation}
\begin{array}{l}
 {{F\left( z \right)8R^2 } \mathord{\left/ {\vphantom {{F\left( z
 \right)8R^2 } L}} \right.  \kern-\nulldelimiterspace} L} = \left[ {E_p
 \left( x \right)\left( {{{3\sqrt \pi } \mathord{\left/ {\vphantom {{3\sqrt
 \pi } 4}} \right.  \kern-\nulldelimiterspace} 4}} \right)\sqrt {{{2\pi }
 \mathord{\left/ {\vphantom {{2\pi } x}} \right. 
 \kern-\nulldelimiterspace} x}} } \right] \\
 \,\,\,\,\,\,\,\,\,\,\,\,\,\,\,\,\,\, \times \left[ {\frac{{6\left( {x + 1}
 \right)^2 \tan ^{ - 1} \left( {{1 \mathord{\left/ {\vphantom {1 {\sqrt
 {x^2 + 2x} }}} \right.  \kern-\nulldelimiterspace} {\sqrt {x^2 + 2x} }}}
 \right) + 3\pi \left( {x + 1} \right)^2 + 2\sqrt {x^2 + 2x} \left( {2x^2 +
 4x + 3} \right)}}{{\left( {{{3\sqrt \pi } \mathord{\left/ {\vphantom
 {{3\sqrt \pi } 4}} \right.  \kern-\nulldelimiterspace} 4}} \right)\sqrt
 {{{2\pi } \mathord{\left/ {\vphantom {{2\pi } x}} \right. 
 \kern-\nulldelimiterspace} x}} \left( {x + 1} \right)\left( {x + 2}
 \right)^2 \sqrt {x^2 + 2x} }}} \right],
\end{array}
\label{CylCyl}
\end{equation}
where now the variable $x=z/2R$. Note that the right hand side of this 
equation is identical to the one in Eq. (\ref{CylSub}).

\subsection{Standing cylinder}

For a cylinder standing upright above a substrate, the proximity force
approximation gives the potential as the bottom area of the cylinder times
the interaction energy per unit area between two halfspaces.  The full PFA gives the same result. For a cylinder of length $L$ with
circular cross section of radius $R$, we have in the non-retarded case
$V\left( z\right) = \pi R^2 E_p \left( z \right)$ and
%43
\begin{equation}
F\left( z \right) = - \frac{{dV\left( z \right)}}{{dz}} = 2\pi R^2
\frac{1}{z}E_p \left( z \right).
\end{equation}
The universal expression is
%44
\begin{equation}
RF = {{2\pi E_p \left( x \right)} \mathord{\left/
 {\vphantom {{2\pi E_p \left( x \right)} x}} \right.
 \kern-\nulldelimiterspace} x},
\end{equation}
 where $x = {z \mathord{\left/ {\vphantom {z R}} \right. 
 \kern-\nulldelimiterspace} R}$.

\section{\label{Cube} Cube-substrate interaction}

\subsection{Straight cube}
For a cube standing upright above a substrate, the proximity force
approximation gives the potential as the bottom area of the cube times
the interaction energy per unit area between two halfspaces.  The full PFA gives the same result. For a cube of side length $L$ we
have in the non-retarded case $V\left( z\right) = L^2 E_p \left( z
\right)$ and
%45
\begin{equation}
F\left( z \right) = - \frac{{dV\left( z \right)}}{{dz}} = 2L^2
\frac{1}{z}E_p \left( z \right).
\end{equation}
The universal expression is
%46
\begin{equation}
LF = {{2 E_p \left( x \right)} \mathord{\left/
 {\vphantom {{2 E_p \left( x \right)} x}} \right.
 \kern-\nulldelimiterspace} x},
\end{equation}
 where $x = {z \mathord{\left/ {\vphantom {z L}} \right. 
 \kern-\nulldelimiterspace} L}$.
 \begin{figure}
\includegraphics[width=10cm]{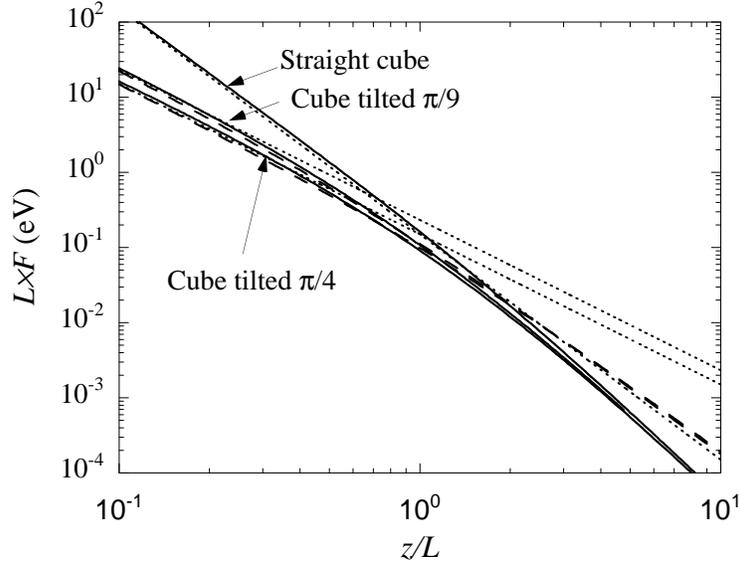}
\caption{The force on a gold cube of side length $L$ above a
gold substrate.  The dotted curves and dashed curves are for the PFA and
full PFA, respectively.  The thin solid curves is the full result from
calculations based on integral equation techniques.}
\label{figu12}
\end{figure}

\subsection{Tilted cube}

Here we tilt the cube the angle $\alpha$ while keeping one of the edges 
parallel to the substrate. The potential is
%47
\begin{equation}
\begin{array}{l}
 V\left( z \right) = \int\limits_S {dS} E_p \left( w \right) =
 \int\limits_0^{L\cos \alpha } {dx} LE_p \left( {z + x\tan \alpha } \right)
 + \int\limits_0^{L\sin \alpha } {dx} LE_p \left( {z + x\cot \alpha }
 \right) \\ \,\,\,\,\,\,\,\,\,\,\,\,\,\,\, = L\cot \alpha \int\limits_z^{z +
 L\sin \alpha } {dw} E_p \left( w \right) + L\tan \alpha \int\limits_z^{z +
 L\cos \alpha } {dw} E_p \left( w \right),
\end{array}
\end{equation}
and the force
%48
\begin{equation}
\begin{array}{l}
 F\left( z \right) = - \frac{{dV\left( z \right)}}{{dz}} = - L\left[ {\cot
 \alpha E_p \left( {z + L\sin \alpha } \right) - \cot \alpha E_p \left( z
 \right)} \right.  \\ \,\,\,\,\,\,\,\,\,\,\,\,\,\,\,\,\,\, + \left.  {\tan
 \alpha E_p \left( {z + L\cos \alpha } \right) - \tan \alpha E_p \left( z
 \right)} \right] \\ \,\,\,\,\,\,\,\,\,\,\,\,\,\, = \frac{{ - L}}{{\sin
 \alpha \cos \alpha }}\left[ {\cos ^2 \alpha E_p \left( {z + L\sin \alpha }
 \right) + \sin ^2 \alpha E_p \left( {z + L\cos \alpha } \right) - E_p
 \left( z \right)} \right].
\end{array}
\end{equation}
The universal expression in the non-retarded limit is
%49
\begin{equation}
LF = \left[ {\frac{1}{{\sin \alpha \cos \alpha }}E_p \left( x \right)}
\right]\left[ {1 - \cos ^2 \alpha \left( {1 + \frac{{\sin \alpha }}{x}}
\right)^{ - 2} - \sin ^2 \alpha \left( {1 + \frac{{\cos \alpha }}{x}}
\right)^{ - 2} } \right],
\end{equation}

where $x=z/L£$. The first factor is the PFA result and the second the 
correction factor due to the full geometry. Note that the $x$ dependence 
of the PFA result changes abruptly when the cube is tilted. In Fig. 
\ref{figu12} we show the results for a gold cube above a gold substrate. 
 The results are for a straight cube and for a cube tilted by ${\pi
 \mathord{\left/ {\vphantom {\pi 9}} \right.  \kern-\nulldelimiterspace}
 9}$ and ${\pi \mathord{\left/ {\vphantom {\pi 4}} \right. 
 \kern-\nulldelimiterspace} 4}$, respectively.  The dotted curves are the
 PFA result and the dashed curves the full PFA result.  The results from
 the integral equation method are represented by thin solid curves.  For the
 straight cube there are no corrections to PFA. For large separations the
 force for both straight and tilted cubes varies as $ F \sim {{L^3 }
 \mathord{\left/ {\vphantom {{L^3 } {z^4 }}} \right. 
 \kern-\nulldelimiterspace} {z^4 }}$; for small separations the force on
 the straight cube varies as $ F \sim {{L^2 } \mathord{\left/ {\vphantom
 {{L^2 } {z^3 }}} \right.  \kern-\nulldelimiterspace} {z^3 }}$; for small
 separations the force on a tilted cube varies as $ F \sim {L
 \mathord{\left/ {\vphantom {L {z^2 }}} \right.  \kern-\nulldelimiterspace}
 {z^2 }}$.  This means that the force at large separations is proportional
 to the volume of the object; at small separations it is proportional to
 the projected area for the straight cube and to the length of the closest
 edge for the tilted cube.

\section{\label{Cone}Cone-substrate interaction}
\begin{figure}
\includegraphics[width=10cm]{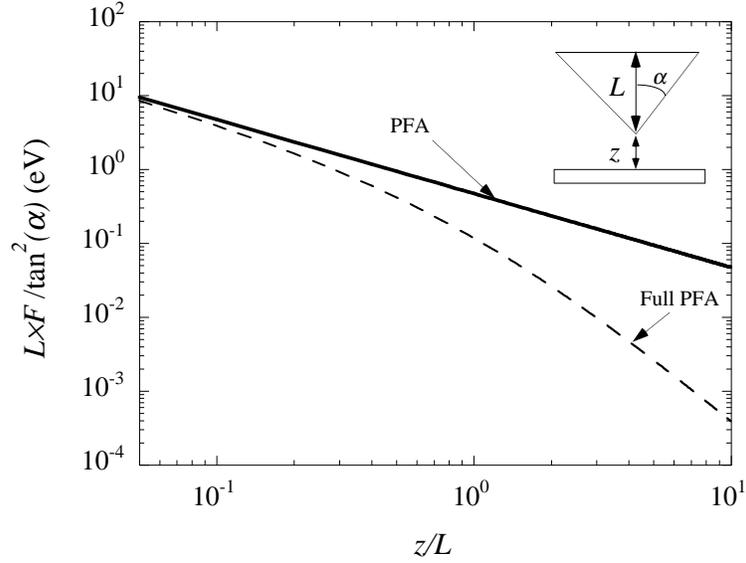}
\caption{The force on a gold cone of height $L$ above a
gold substrate.  Plotted in this way the PFA and full PFA results (thick solid and dashed
curve, respectively) produce universal curves independent of scale and aperture angle
($2\alpha)$.}
\label{figu13}
\end{figure}
\subsection{\label{PointedCone}Cone with a pointed tip}
Let the cone have circular bottom of radius $R$ and be of height $L$. It has its
point towards the substrate at the distance $z$. In this case Eqs. 
(\ref{PFAterms}) and (\ref{PFAFactors}) can not be used. Instead, we write
%50
\begin{equation}
 V\left( h \right) = \int\limits_S {dS} E_p \left( w \right) =
 \int\limits_0^R {dr2\pi r} E_p \left( {h + r{L \mathord{\left/ {\vphantom
 {L R}} \right.  \kern-\nulldelimiterspace} R}} \right),
\end{equation}
and if Eq. (\ref{Evariation}) holds we have
%51
\begin{equation}
\begin{array}{l}
 F\left( z \right) = - \frac{{dV\left( z \right)}}{{dz}} = -
 \int\limits_0^R {dr} \frac{{2\pi rCn}}{{\left( {z + r{L \mathord{\left/
 {\vphantom {L R}} \right.  \kern-\nulldelimiterspace} R}} \right)^{n + 1}
 }} = - 2\pi CnR^2 \int\limits_0^1 {dr} \frac{r}{{\left( {z + rL}
 \right)^{n + 1} }} \\ \,\,\,\,\,\,\,\,\,\,\,\,\, = - \frac{{2\pi CnR^2
 }}{{L^{n + 1} }}\int\limits_0^1 {dr} \frac{r}{{\left( {{z \mathord{\left/
 {\vphantom {z L}} \right.  \kern-\nulldelimiterspace} L} + r} \right)^{n +
 1} }} \\ \,\,\,\,\,\,\,\,\,\,\,\,\, = \frac{{2\pi CR^2 }}{{L^{n + 1} }}
 \times \left\{ \begin{array}{l} \frac{1}{{n - 1}}\left[ {{{\left( {{z
 \mathord{\left/ {\vphantom {z L}} \right.  \kern-\nulldelimiterspace} L} +
 n} \right)} \mathord{\left/ {\vphantom {{\left( {{z \mathord{\left/
 {\vphantom {z L}} \right.  \kern-\nulldelimiterspace} L} + n} \right)}
 {\left( {{z \mathord{\left/ {\vphantom {z L}} \right. 
 \kern-\nulldelimiterspace} L} + 1} \right)^n }}} \right. 
 \kern-\nulldelimiterspace} {\left( {{z \mathord{\left/ {\vphantom {z L}}
 \right.  \kern-\nulldelimiterspace} L} + 1} \right)^n }} - {1
 \mathord{\left/ {\vphantom {1 {\left( {{z \mathord{\left/ {\vphantom {z
 L}} \right.  \kern-\nulldelimiterspace} L}} \right)^{n - 1} }}} \right. 
 \kern-\nulldelimiterspace} {\left( {{z \mathord{\left/ {\vphantom {z L}}
 \right.  \kern-\nulldelimiterspace} L}} \right)^{n - 1} }}} \right];\quad
 n \ne 1 \\ 1 + \ln \left[ {{{\left( {{z \mathord{\left/ {\vphantom {z L}}
 \right.  \kern-\nulldelimiterspace} L}} \right)} \mathord{\left/
 {\vphantom {{\left( {{z \mathord{\left/ {\vphantom {z L}} \right. 
 \kern-\nulldelimiterspace} L}} \right)} {\left( {{z \mathord{\left/
 {\vphantom {z L}} \right.  \kern-\nulldelimiterspace} L} + 1} \right)}}}
 \right.  \kern-\nulldelimiterspace} {\left( {{z \mathord{\left/ {\vphantom
 {z L}} \right.  \kern-\nulldelimiterspace} L} + 1} \right)}}} \right] -
 {{\left( {{z \mathord{\left/ {\vphantom {z L}} \right. 
 \kern-\nulldelimiterspace} L}} \right)} \mathord{\left/ {\vphantom
 {{\left( {{z \mathord{\left/ {\vphantom {z L}} \right. 
 \kern-\nulldelimiterspace} L}} \right)} {\left( {{z \mathord{\left/
 {\vphantom {z L}} \right.  \kern-\nulldelimiterspace} L} + 1} \right)}}}
 \right.  \kern-\nulldelimiterspace} {\left( {{z \mathord{\left/ {\vphantom
 {z L}} \right.  \kern-\nulldelimiterspace} L} + 1} \right)}};\quad n = 1.
 \\ \end{array} \right.  \\ \end{array}
\end{equation}
In our case we have $n=2$ and may write on universal form 
%52
\begin{equation}
F\left( z \right)L = {{2\pi E_p \left( x \right)x^2 \tan ^2 \left( \alpha
\right)} \mathord{\left/ {\vphantom {{2\pi E_p \left( x \right)x^2 \tan ^2
\left( \alpha \right)} {\left[ {x\left( {1 + x} \right)^2 } \right]}}}
\right.  \kern-\nulldelimiterspace} {\left[ {x\left( {1 + x} \right)^2 }
\right]}};\quad x = {z \mathord{\left/ {\vphantom {z L}} \right. 
\kern-\nulldelimiterspace} L}.
\end{equation}
In Fig. \ref{figu13} we plot $L\times F(z)/tan^{2}(\alpha)$ as function
of $z/L$ for a gold cone above a gold substrate.  In PFA  and full PFA this produces a
universal curve for all sizes and for all aperture angles.  We note that
for large $z/L$ the full PFA curve varies as $(z/L)^{-3}$ which means that the force
varies as $z^{-3}$ and is proportional to the projected area of the cone. 
For small relative separations the curve varies in the limit as
$(z/L)^{-1}$ which means that the force varies as $z^{-1}$ and is
independent of the size of the cone. In this geometry the integral equation method has convergency problems in the small separation limit so we omit to include any results here. This method works much better for cones with spherical tips which is the topic of next section.

\subsection{\label{Conetip}Cone with a spherical tip}
The cone with a pointed tip treated in the previous section is an idealization. In all practical cases the tip is rounded. In this section we let the tip be a part of a sphere. We let $L$ be the height of the truncated cone, and $R$ the radius of
curvature of the spherical tip.  The tip connects to the truncated cone so
that the slope of the surface is continuous at the connection.  The nearest
point is $z$ above the substrate.  Let the angle of aperture be $2\alpha$. 
The force in full PFA consists of two parts, $RF = RF_1 + RF_2 $, where the first part
comes from the spherical tip
%53
\begin{equation}
RF_1 = 2\pi E_p \left( x \right)\left\{ {1 - \frac{{x^2 \sin \alpha
}}{{\left[ {x + 1 - \sin \alpha } \right]^2 }} - \left[ {x - \frac{{x^2
}}{{x + 1 - \sin \alpha }}} \right]} \right\};\,\,x = {z \mathord{\left/
{\vphantom {z R}} \right.  \kern-\nulldelimiterspace} R},
\end{equation}
and the second from the truncated cone
%54
\begin{equation}
RF_2 \left( z \right) =  2\pi \tan ^2 \left( \alpha  \right)E_p \left( x \right)x^2 \left[ {\frac{1}{{x + 1 - \sin \left( \alpha  \right)}} - \frac{{x + 1 - \sin \left( \alpha  \right) + 2{L \mathord{\left/
 {\vphantom {L R}} \right.
 \kern-\nulldelimiterspace} R}}}{{\left( {x + 1 - \sin \left( \alpha  \right) + {L \mathord{\left/
 {\vphantom {L R}} \right.
 \kern-\nulldelimiterspace} R}} \right)^2 }}} \right];\quad x = {z \mathord{\left/
 {\vphantom {z R}} \right.
 \kern-\nulldelimiterspace} R},
\end{equation}
where we have assumed that Eq. (\ref{Evariation}) holds with $2$ as an
exponent.

\begin{figure}
\begin{minipage}{20pc}
\vspace{0.5pc}
\includegraphics[width=8cm]{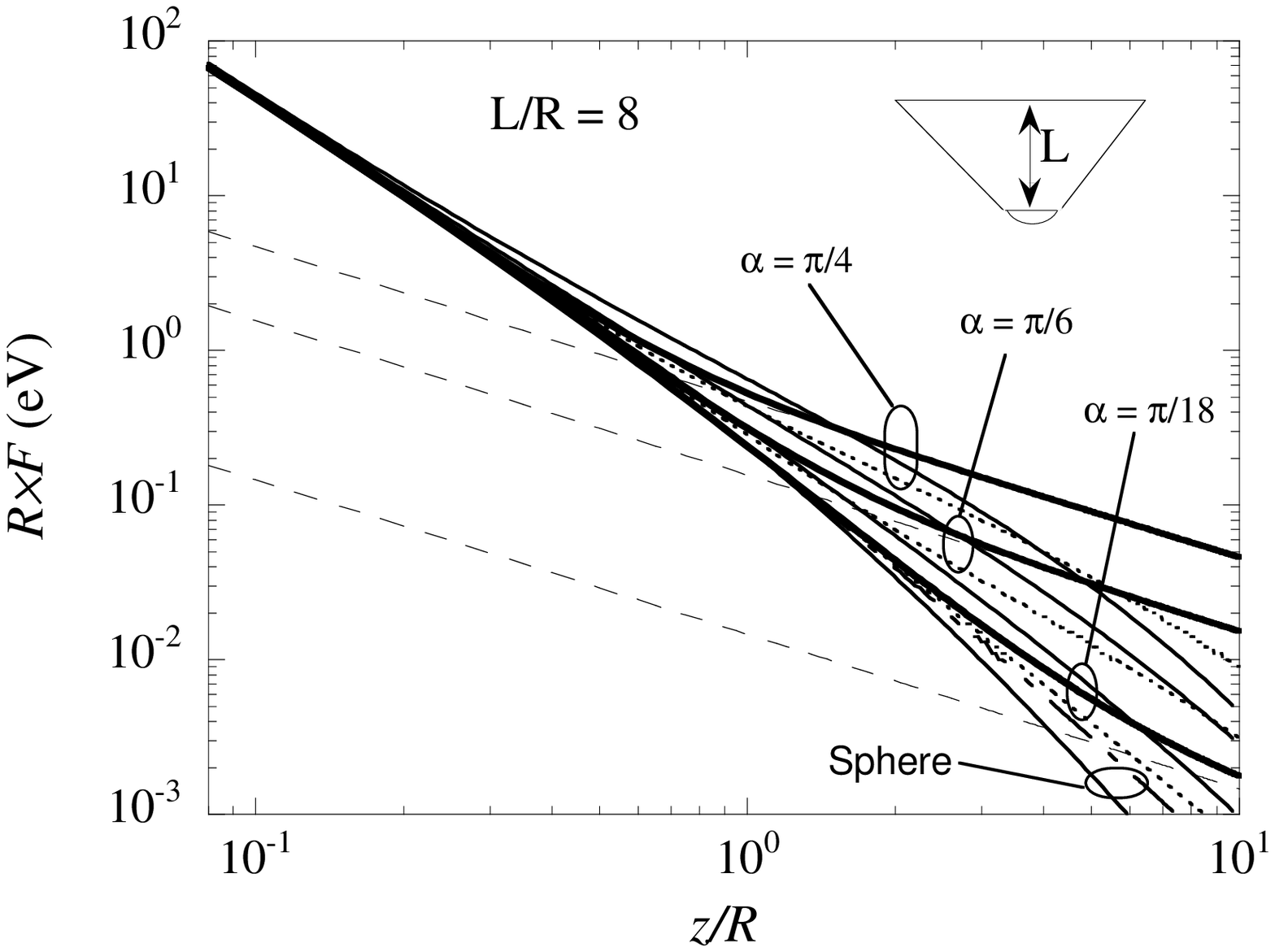}
\caption{The force on a gold cone of height $L$ with a spherical tip with 
radius of curvature $R$ above a gold substrate.  The results are for $L/R =
8$.  The thin solid (dotted) curves are from the full (full PFA)
calculation.  The thick solid curves are the full PFA limit when $L/R \to 
\infty$ ; the dashed curves are corresponding asymptotes. }
\label{figu14}
\end{minipage}\hspace{1pc}
\begin{minipage}{20pc}
\includegraphics[width=8cm]{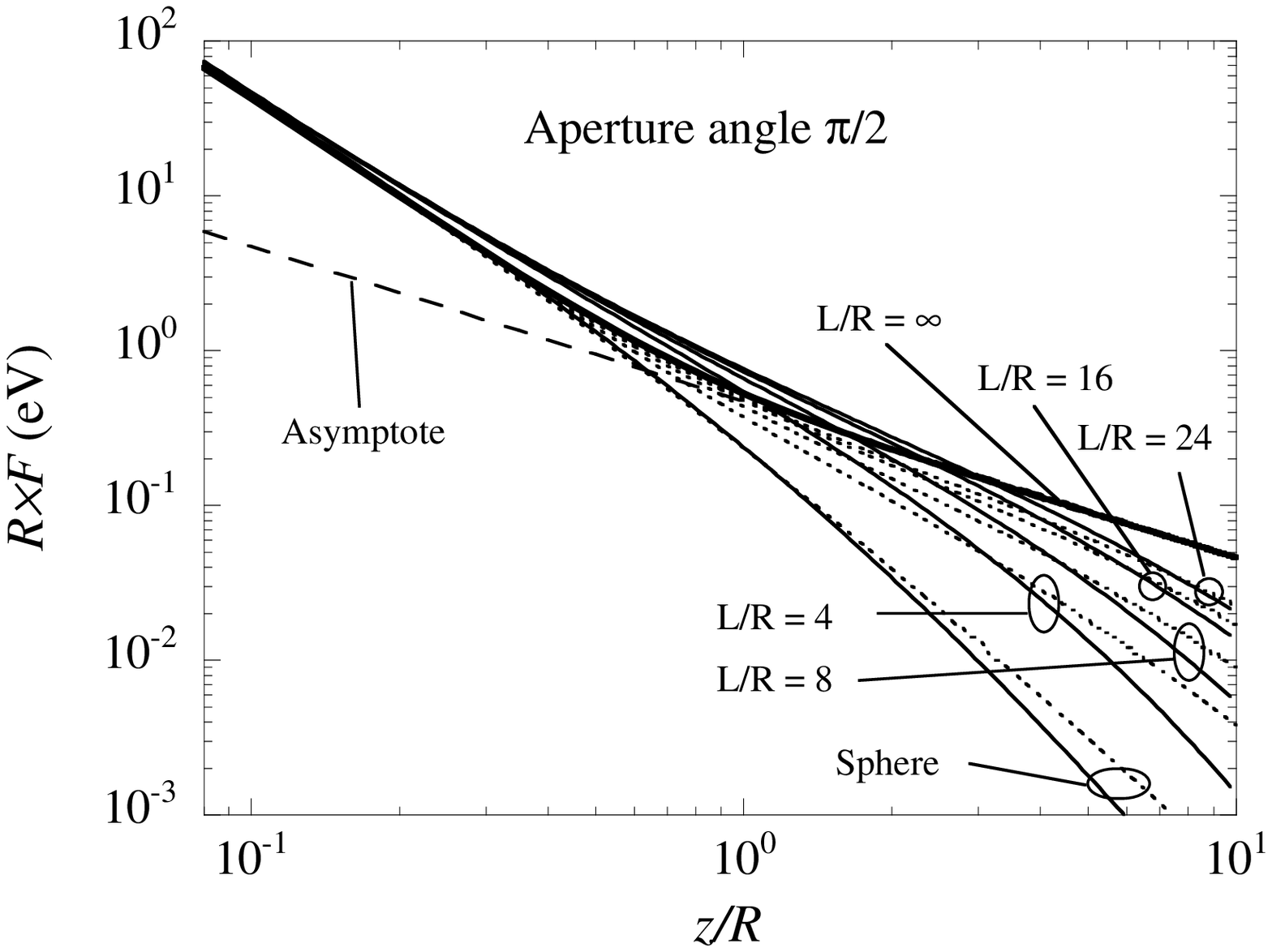}
\caption{Same as Fig.  \ref{figu14} but here the angle of aperture is kept
at the fixed value $\pi /2$ and $L/R$ varies.  The thick solid curve is the
full PFA result in the limit when $L/R \to
\infty$; the dashed curve is the corresponding asymptote.  The thin solid
(dotted) curves are from the full (full PFA) calculation.  }
\label{figu15}
\end{minipage}
\end{figure}
 In Fig.  \ref{figu14} we compare the full PFA result (dotted curves) with
 that from the full calculation (thin solid curves) of the force on a gold
 cone with spherical tip above a gold substrate.  In all cases $L$ is kept
 8 times the radius of curvature, $R$, of the tip.  The thick solid curves
 are the limiting result of the full PFA when the height of the cone goes
 toward infinity; the dashed curves are the corresponding asymptotes, $ RF
 = 2\pi \tan ^2 \left( \alpha \right)E_p \left( x \right)x $.  All set of
 curves are for the $\alpha$ values ${\pi \mathord{\left/ {\vphantom {\pi
 4}} \right.  \kern-\nulldelimiterspace} 4}$, ${\pi \mathord{\left/
 {\vphantom {\pi 6}} \right.  \kern-\nulldelimiterspace} 6}$ and ${\pi
 \mathord{\left/ {\vphantom {\pi 18}} \right.  \kern-\nulldelimiterspace}
 18}$, respectively.
In Fig. \ref{figu15} we keep the aperture angle fix with the value ${\pi
 \mathord{\left/ {\vphantom {\pi 2}} \right.  \kern-\nulldelimiterspace}
 2}$ and vary $L/R$.  The thin solid (dotted) curves
are from the full (full PFA) calculation.  The thick solid curve is the
full PFA result in the limit when $L/R \to \infty$; the dashed curve is 
the corresponding asymptote. We see that the full PFA results stay below the 
thick solid curve for all separations and comes closer the larger the $L/R$ 
value. The results from the full non-retarded calculation cross the 
thick solid curve.

\section{\label{Wings}Wings-substrate interaction}

\begin{figure}
\includegraphics[width=10cm]{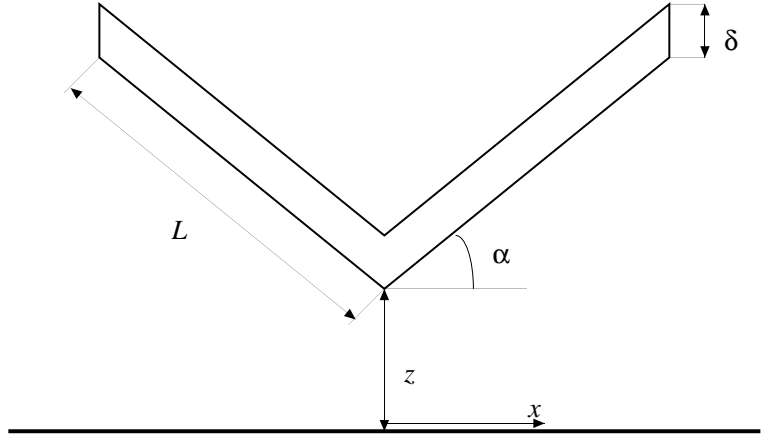}
\caption{Definition of the object wings. }
\label{figu16}
\end{figure}

In this section we consider a geometrical object, which we call wings,
described in Fig.  \ref{figu16}, above a substrate.  The object has the
extension $L$ in the direction perpendicular to the plane of the figure. 
Thus, the bottom and top surfaces are squares of side length $L$. The 
thickness of the wings is $\delta$. 

\subsection{\label{ThickWings}Thick wings}

\begin{figure}
\begin{minipage}{20pc}
\includegraphics[width=8cm]{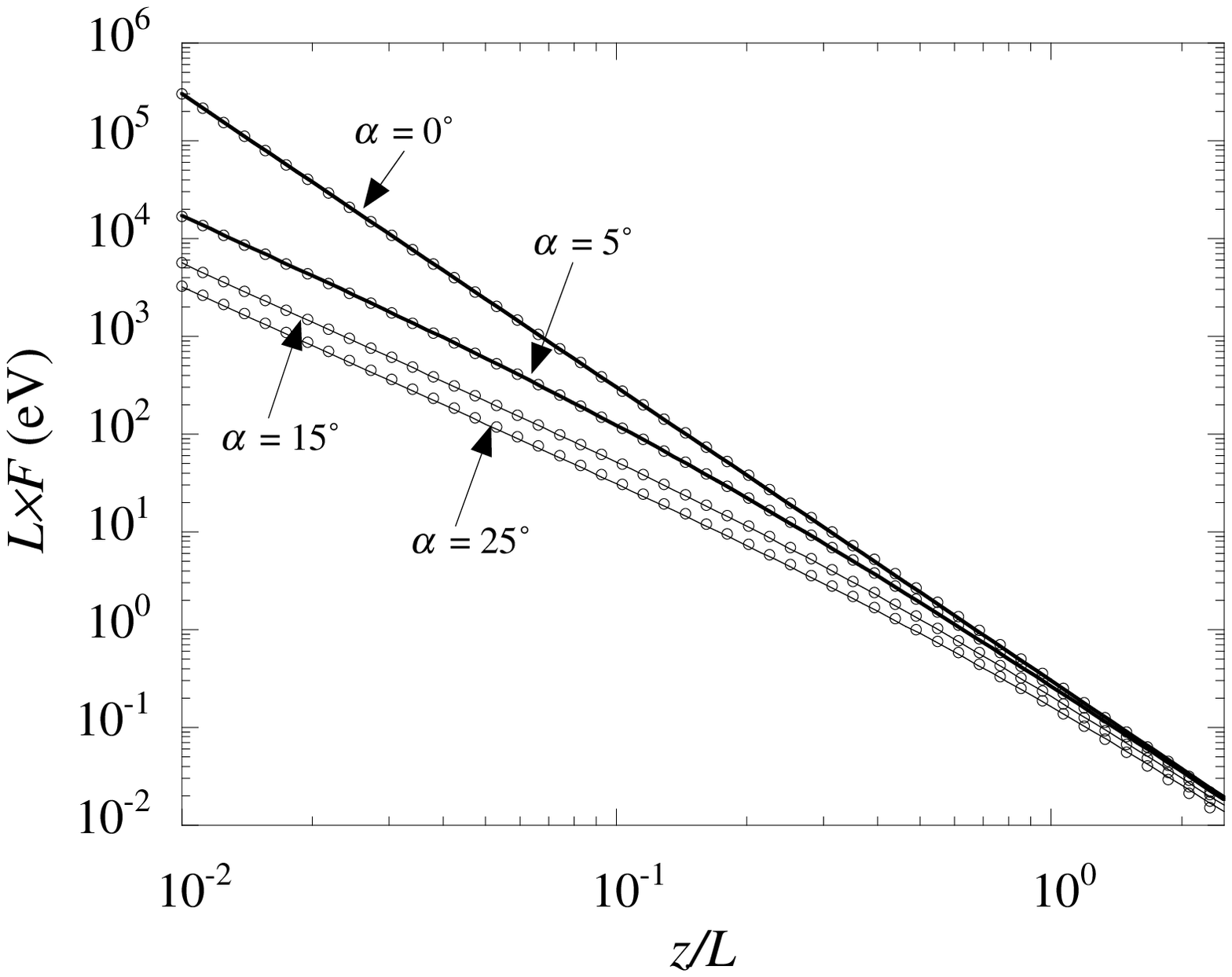}
\caption{The force on thick gold wings above a gold substrate. The lines 
is the full PFA result and the circles is the full result from a 2D version of the integral equation method. Note that the slope of the zero angle curve is 
different from that of the others at the small $x$ end of the figure}
\label{figu17}
\end{minipage}\hspace{1pc}
\begin{minipage}{20pc}
\vspace{3.5pc}
\includegraphics[width=8cm]{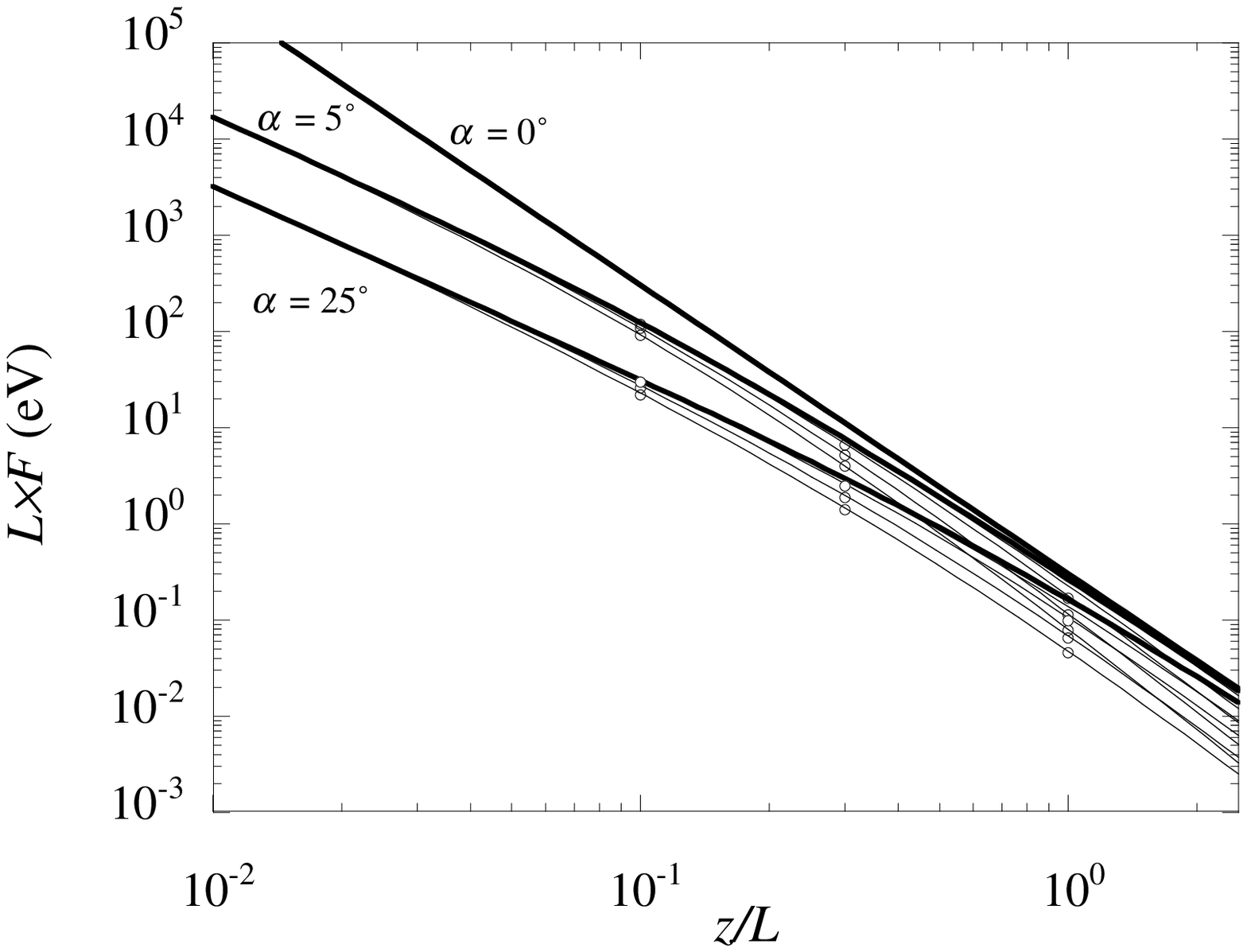}
\vspace{1pc}
\caption{The force on goldwings above a gold substrate. The results are for 
angles $0^\circ $, $5^\circ $, and $25^\circ $.  Thick curves (thin curves)
are full PFA results (full results).  The full results are for thickness $5\% $,
$10\% $, $25\% $, $50\% $, and $100\%$.  The thicker the curve the closer
it agrees with the full PFA result for thick wings.  The $100\%$ curve has completely merged
with the corresponding full PFA curve for thick wings in the whole $z/L$ range covered in the
figure.  The full PFA results for thickness $5\% $, $10\% $, and $25\% $
are indicated by circles.}
\label{figu18}
\end{minipage}
\end{figure}

In the standard PFA the result is independent of the thickness and is the
result for infinite thickness.  The potential is in full PFA found as
%55
\begin{equation}
\begin{array}{l}
 V\left( z \right) = \int\limits_S {dS} E_p \left( w \right) =
 2\int\limits_0^{L\cos \alpha } {dx} LE_p \left( {z + x\tan \alpha }
 \right) \\ \,\,\,\,\,\,\,\,\,\,\,\,\, = 2L\cot \alpha \int\limits_z^{z +
 L\sin \alpha } {dw} E_p \left( w \right).
\end{array}
\end{equation}
and the force as
%56
\begin{equation}
\begin{array}{l}
 F\left( z \right) = - \frac{{dV\left( z \right)}}{{dz}} = 2L\left[ {\cot
 \alpha E_p \left( {z + L\sin \alpha } \right) - \cot \alpha E_p \left( z
 \right)} \right] \\ \,\,\,\,\,\,\,\,\,\,\,\,\,\, = 2L\cot \alpha \left[
 {E_p \left( {z + L\sin \alpha } \right) - E_p \left( z \right)} \right]
 \end{array}
\end{equation}
 In the non-retarded version the result is
%57
\begin{equation}
\begin{array}{l}
 LF\left( z \right) = 2L^2 \cot \alpha E_p \left( z \right)\left[
 {\frac{{z^2 }}{{\left( {z + L\sin \alpha } \right)^2 }} - 1} \right] \\
 \,\,\,\,\,\,\,\,\,\,\,\,\,\,\,\,\,\, = - 4\cos \alpha E_p \left( x
 \right)\frac{1}{x}\left[ {\frac{{\left( {1 + {{\sin \alpha }
 \mathord{\left/ {\vphantom {{\sin \alpha } {2x}}} \right. 
 \kern-\nulldelimiterspace} {2x}}} \right)}}{{\left( {1 + {{\sin \alpha }
 \mathord{\left/ {\vphantom {{\sin \alpha } x}} \right. 
 \kern-\nulldelimiterspace} x}} \right)^2 }}} \right] \\
 \,\,\,\,\,\,\,\,\,\,\,\,\,\,\,\,\, = - 2\cot \alpha E_p \left( x
 \right)\left[ {\frac{{\left( {1 + {{2x} \mathord{\left/ {\vphantom {{2x}
 {\sin \alpha }}} \right.  \kern-\nulldelimiterspace} {\sin \alpha }}}
 \right)}}{{\left( {1 + {x \mathord{\left/ {\vphantom {x {\sin \alpha }}}
 \right.  \kern-\nulldelimiterspace} {\sin \alpha }}} \right)^2 }}}
 \right];\quad x = {z \mathord{\left/ {\vphantom {z L}} \right. 
 \kern-\nulldelimiterspace} L},
\end{array}
\end{equation}
where the second line is useful in finding the zero angle limit,$ - 4{{E_p
\left( x \right)} \mathord{\left/ {\vphantom {{E_p \left( x \right)} x}}
\right.  \kern-\nulldelimiterspace} x}$ , or the large $x$ asymptote, $ -
4\cos \alpha {{E_p \left( x \right)} \mathord{\left/ {\vphantom {{E_p
\left( x \right)} x}} \right.  \kern-\nulldelimiterspace} x}$ ; the third
line to find the small $x$ asymptote, $ - 2\cot \alpha E_p \left( x
\right)$. In Fig.  \ref{figu17} we give the results for gold wings above a 
gold substrate.  The solid curves are the full PFA results for the angles 0, 5,
15, and 25 degrees.  The circles are the corresponding results from the
full integral equation calculation.  In this calculation one has to choose a big, but finite
thickness.  We have chosen it to be equal to $L$.  We have used a 2D version of the theory, i.e., we have assumed that the wings have
an infinite extention in the direction perpendicular to the plane of the
figure.  We obtain the force per unit length and multiply this with the
actual length $L$.  One may alternatively perform the calculation for the
actual length $L$ directly, i.e., use a 3D version of the theory.  These two results are expected to be different
for large $x$ values; in full PFA there is no difference.  Note that the full PFA
works very well for this problem.  For non-zero angles the force is
proportional to the length of the edge at small distances and to the
projected surface for large distances; for zero angle it is proportional to 
the projected surface area at all distances.

\subsection{\label{ThinWings}Thin wings}

For wings of general thickness one finds in full PFA
%58
\begin{equation}
\begin{array}{l}
 F\left( z \right) = - \frac{{dV\left( z \right)}}{{dz}} = 2L\cot \alpha
 \left[ {E_p \left( {z + L\sin \alpha ,\delta } \right) - E_p \left(
 {z,\delta } \right)} \right] \\ \,\,\,\,\,\,\,\,\,\,\,\,\,\, = 2L\cot
 \alpha \frac{1}{{L^2 }}\left[ {E_p \left( {x + \sin \alpha ,{\delta
 \mathord{\left/ {\vphantom {\delta L}} \right.  \kern-\nulldelimiterspace}
 L}} \right) - E_p \left( {x,{\delta \mathord{\left/ {\vphantom {\delta L}}
 \right.  \kern-\nulldelimiterspace} L}} \right)} \right];\quad x = {z
 \mathord{\left/ {\vphantom {z L}} \right.  \kern-\nulldelimiterspace} L},
 \end{array}
\end{equation}
and
%59
\begin{equation}
 LF\left( {x,{\delta \mathord{\left/ {\vphantom {\delta L}} \right. 
 \kern-\nulldelimiterspace} L}} \right) = 2\cot \alpha \left[ {E_p \left(
 {x + \sin \alpha ,{\delta \mathord{\left/ {\vphantom {\delta L}} \right. 
 \kern-\nulldelimiterspace} L}} \right) - E_p \left( {x,{\delta
 \mathord{\left/ {\vphantom {\delta L}} \right.  \kern-\nulldelimiterspace}
 L}} \right)} \right].
\end{equation}

In Fig.  \ref{figu18} the full PFA result for thick wings of angles $0^\circ $,
$5^\circ $, and $25^\circ $ is shown as thick curves.  The thin curves
represent he full result for thickness $5\% $, $10\% $, $25\% $, $50\% $,
and $100\%$.  The thicker the curve the closer it agrees with the full PFA
result for thick wings.  The $100\%$ curve has completely merged with the corresponding full PFA
curve for thick wings in the whole $x$ range covered in the figure.  The full PFA results
for thickness $5\% $, $10\% $, and $25\% $ are indicated by circles.  We
find that the results from the full PFA even here agree quite well with the
results from the more well founded result from the 2D version of the integral equation method.

If the thickness becomes extreemely small the wings behave as a bent 
two-dimensional metallic film,
%60
\begin{equation}
\begin{array}{l}
 LF\left( z \right) = 2L^2 \cot \alpha \left[ {E_p \left( {z + L\sin \alpha
 } \right) - E_p \left( z \right)} \right] \\
 \,\,\,\,\,\,\,\,\,\,\,\,\,\,\,\,\,\, \approx 2L^2 \cot \alpha 0.02477\sqrt
 {{{n\hbar ^2 e^2 } \mathord{\left/ {\vphantom {{n\hbar ^2 e^2 } {m_e }}}
 \right.  \kern-\nulldelimiterspace} {m_e }}} \sqrt \delta \left[ {z^{ - {5
 \mathord{\left/ {\vphantom {5 2}} \right.  \kern-\nulldelimiterspace} 2}}
 - \left( {z + L\sin \alpha } \right)^{ - {5 \mathord{\left/ {\vphantom {5
 2}} \right.  \kern-\nulldelimiterspace} 2}} } \right] \\
 \,\,\,\,\,\,\,\,\,\,\,\,\,\,\,\,\,\, = 0.04954\cot \alpha \sqrt {{{n\hbar
 ^2 e^2 } \mathord{\left/ {\vphantom {{n\hbar ^2 e^2 } {m_e }}} \right. 
 \kern-\nulldelimiterspace} {m_e }}} \sqrt {{\delta \mathord{\left/
 {\vphantom {\delta L}} \right.  \kern-\nulldelimiterspace} L}} \left[ {x^{
 - {5 \mathord{\left/ {\vphantom {5 2}} \right.  \kern-\nulldelimiterspace}
 2}} - \left( {x + \sin \alpha } \right)^{ - {5 \mathord{\left/ {\vphantom
 {5 2}} \right.  \kern-\nulldelimiterspace} 2}} } \right] \\
 \,\,\,\,\,\,\,\,\,\,\,\,\,\,\,\,\, = \left| {\alpha \to {\rm{0}}} \right|
 = 0.12385\sqrt {{{n\hbar ^2 e^2 } \mathord{\left/ {\vphantom {{n\hbar ^2
 e^2 } {m_e }}} \right.  \kern-\nulldelimiterspace} {m_e }}} \sqrt {{\delta
 \mathord{\left/ {\vphantom {\delta L}} \right.  \kern-\nulldelimiterspace}
 L}} x^{ - {7 \mathord{\left/ {\vphantom {7 2}} \right. 
 \kern-\nulldelimiterspace} 2}} \\ \,\,\,\,\,\,\,\,\,\,\,\,\,\,\,\,\, =
 \left| {x \to {\rm{0}}} \right| = 0.04954\cot \alpha \sqrt {{{n\hbar ^2
 e^2 } \mathord{\left/ {\vphantom {{n\hbar ^2 e^2 } {m_e }}} \right. 
 \kern-\nulldelimiterspace} {m_e }}} \sqrt {{\delta \mathord{\left/
 {\vphantom {\delta L}} \right.  \kern-\nulldelimiterspace} L}} x^{ - {5
 \mathord{\left/ {\vphantom {5 2}} \right.  \kern-\nulldelimiterspace} 2}}
 \\ \,\,\,\,\,\,\,\,\,\,\,\,\,\,\,\,\, = \left| {x \to \infty } \right| =
 0.12385\cos \alpha \sqrt {{{n\hbar ^2 e^2 } \mathord{\left/ {\vphantom
 {{n\hbar ^2 e^2 } {m_e }}} \right.  \kern-\nulldelimiterspace} {m_e }}}
 \sqrt {{\delta \mathord{\left/ {\vphantom {\delta L}} \right. 
 \kern-\nulldelimiterspace} L}} x^{ - {7 \mathord{\left/ {\vphantom {7 2}}
 \right.  \kern-\nulldelimiterspace} 2}}.  
\end{array}
\end{equation}

\begin{figure}
\includegraphics[width=10cm]{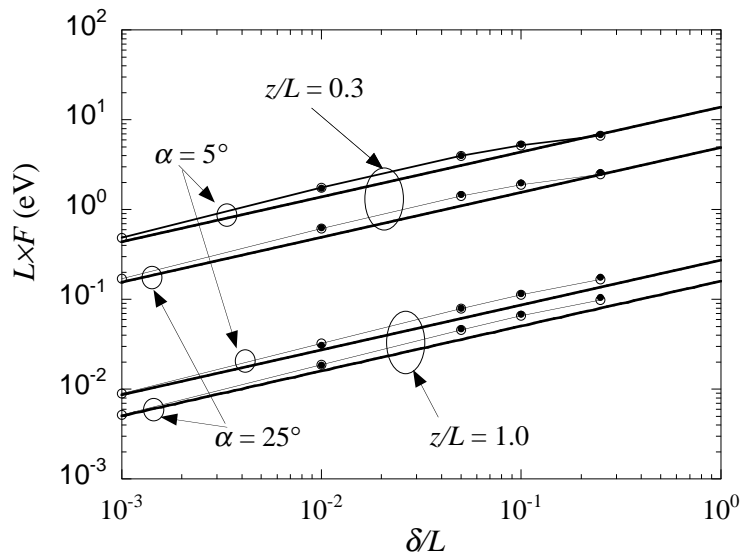}
\caption{The force on gold wings of angles
$5^\circ $ and $25^\circ $, above a gold substrate at two distances 
as function of thickness. The straight lines are the full PFA result for a bent 2D metal film; the open (filled) circles are the full PFA ( 2D version of the integral equation method) results for the actual wing thickness, $\delta$. }
\label{figu19}
\end{figure}

In Fig.  \ref{figu19} we study how the results approach the strictly two 
dimensional  limit (straight line) when the thickness goes towards zero. 
The open circles connected with lines is the result from the full PFA. The 
result from the full calculation using the 2D version of the integral equation method is indicated by filled circles. 

\section{\label{Summary}Summary and conclusions}

In this work we have performed a critical test of the validity of the Proximity Force Approximation. This was done by comparisons with the results from calculation methods of more solid foundation, multi-pole expansions and a method based on an integral equation for the potentials. Calculations were performed for a variety of geometries including edges and corners with different orientations; we studied coated objects and free standing shells.

We extended the traditional PFA in two ways; we did not just keep the small separation asymptote but took the geometrical shapes of the objects fully into account; we took the finite thickness of the coatings and free standing shells into account.

\begin{table*}
\caption{\label{tab:table1}Summary of the PFA result and the geometrical 
correction factor for different geometries. The last column gives the 
lowest order term in the expansion of the correction factor.}
\begin{ruledtabular}
\begin{tabular}{cccc}
Geometry&PFA&$\begin{array}{l}Corr.\,Fact.  \\ = 1 + \epsilon + \ldots \\
\end{array}$ &$\epsilon$ \\
\hline
 sphere-substrate&$R \times F = 2\pi E_p \left( x \right);\;x = {z \mathord{\left/
 {\vphantom {z R}} \right.  \kern-\nulldelimiterspace} R}$&$ {1
 \mathord{\left/ {\vphantom {1 {\left( {x + 1} \right)}}} \right. 
 \kern-\nulldelimiterspace} {\left( {x + 1} \right)}}$ &$-x$\\
 
sphere-sphere&$8R\times F = 2\pi E_p \left( x \right);\;x = {z \mathord{\left/
 {\vphantom {z {2R}}} \right.  \kern-\nulldelimiterspace} {2R}}$&$ {1
 \mathord{\left/ {\vphantom {1 {\left( {x + 1} \right)}}} \right. 
 \kern-\nulldelimiterspace} {\left( {x + 1} \right)}}$ &$-x$\\
 
oblate-substrate&$B\gamma^{ - 2} \times F = 2\pi E_p \left( x \right);\;x = {z
 \mathord{\left/ {\vphantom {z B}} \right.  \kern-\nulldelimiterspace}
 B}$&$ {1 \mathord{\left/ {\vphantom {1 {\left( {x + 1} \right)}}} \right. 
 \kern-\nulldelimiterspace} {\left( {x + 1} \right)}}$ &$-x$\\ 

prolate-substrate&$A\gamma
 ^2 \times F = 2\pi E_p \left( x \right);\;x = {z \mathord{\left/
 {\vphantom {z A}} \right.  \kern-\nulldelimiterspace} A}$&$ {1
 \mathord{\left/ {\vphantom {1 {\left( {x + 1} \right)}}} \right. 
 \kern-\nulldelimiterspace} {\left( {x + 1} \right)}}$&$-x$\\
 
oblate-oblate&$8B\gamma ^{ - 2} \times F = 2\pi E_p \left( x \right);\;x = {z
 \mathord{\left/ {\vphantom {z {2B}}} \right.  \kern-\nulldelimiterspace}
 {2B}}$&$ {1 \mathord{\left/ {\vphantom {1 {\left( {x + 1} \right)}}}
 \right.  \kern-\nulldelimiterspace} {\left( {x + 1} \right)}}$ &$-x$\\
 
prolate-prolate&$8A\gamma ^2 \times F = 2\pi E_p \left( x \right);\;x = {z
 \mathord{\left/ {\vphantom {z {2A}}} \right.  \kern-\nulldelimiterspace}
 {2A}}$&$ {1 \mathord{\left/ {\vphantom {1 {\left( {x + 1} \right)}}}
 \right.  \kern-\nulldelimiterspace} {\left( {x + 1} \right)}}$ &$-x$\\

 cylinder-substrate&$R^2 L^{ - 1} \times F = \left( {{{3\pi } \mathord{\left/ {\vphantom
 {{3\pi } 4}} \right.  \kern-\nulldelimiterspace} 4}} \right)\sqrt {{2
 \mathord{\left/ {\vphantom {2 x}} \right.  \kern-\nulldelimiterspace} x}}
 E_p \left( x \right);\;x = {z \mathord{\left/ {\vphantom {z R}} \right. 
 \kern-\nulldelimiterspace} R}$&$\footnote{ The correction factor is:
  $\left[ {\frac{{6\left( {x + 1} \right)^2 \tan ^{ - 1} \left( {{1
 \mathord{\left/ {\vphantom {1 {\sqrt {x^2 + 2x} }}} \right. 
 \kern-\nulldelimiterspace} {\sqrt {x^2 + 2x} }}} \right) + 3\pi \left( {x
 + 1} \right)^2 + 2\sqrt {x^2 + 2x} \left( {2x^2 + 4x + 3}
 \right)}}{{\left( {{{3\sqrt \pi } \mathord{\left/ {\vphantom {{3\sqrt \pi
 } 4}} \right.  \kern-\nulldelimiterspace} 4}} \right)\sqrt {{{2\pi }
 \mathord{\left/ {\vphantom {{2\pi } x}} \right. 
 \kern-\nulldelimiterspace} x}} \left( {x + 1} \right)\left( {x + 2}
 \right)^2 \sqrt {x^2 + 2x} }}} \right]$} $ &$-x/4$\\ 

cylinder-cylinder&$8R^2 L^{ - 1}
 \times F = \left( {{{3\pi } \mathord{\left/ {\vphantom {{3\pi } 4}}
 \right.  \kern-\nulldelimiterspace} 4}} \right)\sqrt {{2 \mathord{\left/
 {\vphantom {2 x}} \right.  \kern-\nulldelimiterspace} x}} E_p \left( x
 \right);\;x = {z \mathord{\left/ {\vphantom {z {2R}}} \right. 
 \kern-\nulldelimiterspace} {2R}}$&$\footnotemark[1] $ &$-x/4$\\ 

straight cube-substrate&$L \times F = {{2E_p \left( x \right)} \mathord{\left/
 {\vphantom {{2E_p \left( x \right)} x}} \right.
 \kern-\nulldelimiterspace} x};\;x = {z \mathord{\left/
 {\vphantom {z L}} \right.
 \kern-\nulldelimiterspace} L}$&$1$&$0$\\

tilted cube-substrate&$L \times F = {{E_p \left( x \right)} \mathord{\left/ {\vphantom {{E_p
\left( x \right)} {\left( {\sin \alpha \cos \alpha } \right)}}} \right. 
\kern-\nulldelimiterspace} {\left( {\sin \alpha \cos \alpha } \right)}};\;x
= {z \mathord{\left/ {\vphantom {z L}} \right.  \kern-\nulldelimiterspace}
L}$&$\footnote{The correction factor is:$\left[ {1 -
\cos ^2 \alpha \left( {1 + {{\sin \alpha } \mathord{\left/
 {\vphantom {{\sin \alpha } x}} \right.  \kern-\nulldelimiterspace} x}}
 \right)^{ - 2} - \sin ^2 \alpha \left( {1 + {{\cos \alpha }
 \mathord{\left/ {\vphantom {{\cos \alpha } x}} \right. 
 \kern-\nulldelimiterspace} x}} \right)^{ - 2} }
 \right]$}$&$\footnote{Lowest order correction term is: $ - \left[ {\tan ^2
 \alpha + \cot ^2 \alpha } \right]x^2 $}$\\

pointed cone-substrate&$L\left( {\tan \alpha } \right)^{ - 2} \times F = 2\pi xE_p \left( x
 \right);\;x = {z \mathord{\left/ {\vphantom {z L}} \right. 
 \kern-\nulldelimiterspace} L}$&${1 \mathord{\left/ {\vphantom {1 {\left(
 {x + 1} \right)}}} \right.  \kern-\nulldelimiterspace} {\left( {x + 1}
 \right)}}^2 $ &$-2x$\\
 
thick wings-substrate &$L \times F =  - 2\cot \alpha E_p \left( x \right);\;x = {z \mathord{\left/
 {\vphantom {z L}} \right.
 \kern-\nulldelimiterspace} L}$
&$\frac{{\left( {1 + 2{x \mathord{\left/
 {\vphantom {x {\sin \alpha }}} \right.
 \kern-\nulldelimiterspace} {\sin \alpha }}} \right)}}{{\left( {1 + {x \mathord{\left/
 {\vphantom {x {\sin \alpha }}} \right.
 \kern-\nulldelimiterspace} {\sin \alpha }}} \right)^2 }}$
&$ - \left( {{x \mathord{\left/
 {\vphantom {x {\sin \alpha }}} \right.
 \kern-\nulldelimiterspace} {\sin \alpha }}} \right)^2 $\\

\end{tabular}
\end{ruledtabular}
\end{table*}
Throughout this work we have presented universal figures, independent of 
the system scale. A word of caution is in place. All results are limited to 
the non-retarded range. This means that the range of validity in each 
figure depends on the scaling parameter.  
At the present status of the available measurement techniques one requires big enough objects, in the order of tens of micrometers, in close proximity, in the range of hundreds of nanometers. In a majority of cases this means that one is in the non-retarded limit and that the full PFA should be a good approximation.

We have noted two things in particular.  One is that the force between a
cube and a substrate drastically changes character when the cube goes from
standing straight to being slightly tilted. This also holds for the wings when going from zero angle to finite. The distance dependence of the force changes. This means that in experiments it might be better to keep a small but finite angle to the substrate instead of putting too much effort in trying to get perfect alignment; in the interpretation one should then use the full PFA result valid for the finite angle.

The PFA results and geometrical correction factors for most of the 
geometries studied in this work is summarized in Table \ref{tab:table1}.

\begin{acknowledgments}
This research was sponsored by EU within the EC-contract No:012142-NANOCASE
and support from the VR Linn\'{e} Centre LiLi-NFM and from CTS is 
gratefully acknowledged.
\end{acknowledgments}

 %{99}

\end{document}